\documentclass[useAMS]{mn2e}

\usepackage{graphicx} 
\usepackage{dcolumn}  
\usepackage{bm}       
\usepackage{amssymb,amsmath}  
\usepackage{color}
\usepackage{mnras_cite}  
\usepackage{url}

\def\etal{et al.\ }
\def\Mpc{\, h^{-1} \, {\rm Mpc}}
\def\kpc{\, h^{-1} \, {\rm Kpc}}
\def\Mo{\, h^{-1} \, {\rm M_{\odot}}}

\begin{document}

\title[Are the HOD predictions consistent with large scale galaxy clustering?]{Are the halo occupation predictions consistent with large scale galaxy clustering?}

\author[Arnau Pujol, Enrique Gazta\~{n}aga]{Arnau Pujol\thanks{E-mail: pujol@ice.cat}, Enrique Gazta\~{n}aga\\
Institut de Ci\`{e}ncies de l'Espai (ICE, IEEC/CSIC), E-08193 Bellaterra (Barcelona), Spain\\
}
\date{Accepted xxxx. Received xxx}

\pagerange{\pageref{firstpage}--\pageref{lastpage}} \pubyear{2012}

\maketitle

\label{firstpage}

\begin{abstract}
We study how well we can reconstruct the two-point clustering of galaxies on linear scales, as a function of mass and luminosity, using the halo occupation distribution (HOD) in several semi-analytical models (SAMs) of galaxy formation from the Millennium Simulation. We find that HOD with Friends of Friends groups can reproduce galaxy clustering better than gravitationally bound haloes. This indicates that Friends of Friends groups are more directly related to the clustering of these regions than the bound particles of the overdensities. In general we find that the reconstruction works at best to $\simeq 5\%$ accuracy: it underestimates the bias for bright galaxies. This translates to an overestimation of $50\%$ in the halo mass when we use clustering to calibrate mass. We also found a degeneracy on the mass prediction from the clustering amplitude that affects all the masses. This effect is due to the clustering dependence on the host halo substructure, an indication of assembly bias. We show that the clustering of haloes of a given mass increases with the number of subhaloes, a result that only depends on the underlying matter distribution. As the number of galaxies increases with the number of subhaloes in SAMs, this results in a low bias for the HOD reconstruction. We expect this effect to apply to other models of galaxy formation, including the real Universe, as long  as the number of galaxies increases with the number of subhaloes. We have also found that the reconstructions of galaxy bias from the HOD model fails for low mass haloes with $M \lesssim 3-5 \times 10^{11} \Mo$. We find that this is because galaxy clustering is more strongly affected by assembly bias for these low masses.

\end{abstract}


\maketitle
\section{Introduction}

Understanding the link between galaxies and dark matter is one of the fundamental problems that makes precision cosmology difficult to reach. Nowadays cosmological simulations provide accurate measurements of the dark matter distribution of the Universe, but we need to relate the dark matter to galaxy distributions in order to compare to observations.

There are several empirical models of galaxy formation that allow us to populate dark matter simulations with galaxies. On one side, the Halo Occupation Distribution (HOD) (e.g.\ Berlind \& Weinberg 2002) formalism uses the Halo Model (e.g.\ Cooray \& Sheth 2002) to describe the population of galaxies in haloes according to the properties of the host haloes. In many cases the models of galaxy formation assume that the properties and population of galaxies depend only on the halo mass. The population of galaxies is then described by the probability $P(N|M)$ that a halo of virial mass $M$ contains $N$ galaxies of a given type. One can then calculate galaxy clustering from the combination of the HOD with the clustering of halos if we assume that the clustering of haloes depends only on the halo mass. Sheth \etal (2001) used the GIF simulations (Kauffmann \etal 1999) to model the two-point clustering of galaxies from the clustering of haloes and the HOD. They found a good agreement with semi-analytical models. 

If these assumptions are valid we can use galaxy surveys to obtain the relations between properties of galaxies and halo mass, to measure the clustering of dark matter haloes, as well as halo masses (Zehavi \etal 2005, Zheng, Coil and Zehavi 2007, Zehavi \etal 2011, Coupon \etal 2012, Geach \etal 2012). Skibba \etal (2006) used the assumptions from Zehavi \etal (2005) to study the relation between halo mass and satellite and central luminosities, finding a strong relation between central galaxy luminosities and halo mass, and weak dependence for the satellite galaxies. Moster \etal (2010) assumed the HOD to study the relation between the stellar mass of galaxies and halo mass, finding agreement with galaxy clustering in the Sloan Digital Sky Survey (SDSS). Kravtsov \etal (2004) found a relation between the halo mass dependence of populations of subhaloes and the HOD of baryonic simulations and semi-analytical models of galaxies, an indication that the distribution of galaxies can be closely related to the distribution of subhaloes.

However, some studies indicate that several properties of galaxy and halo clustering depend on properties of dark matter haloes other than mass, such as halo formation time, density concentration or subhalo occupation number (Gao \etal 2005, Wechsler \etal 2006, Croft \etal 2012). Wechsler \etal (2006) also saw that the dependence of halo clustering on halo formation time changes with mass. Some studies suggested the idea of adding a second halo property to the HOD model (Abbas \& Sheth 2005, Wechsler \etal 2006, Tinker, Weinberg \& Warren 2006). Croft \etal (2012) studied the clustering dependence of haloes on the occupation number of subhaloes, and found that for fixed masses, the halo bias depends strongly on the number of subhaloes per halo. They found that, for fixed occupation number of subhaloes, the halo bias depends on mass. They also found a strong anticorrelation between clustering and mass for highly occupied haloes. As galaxies possibly follow subhalo gravitational potentials, this dependence can also be found for galaxies, as we show in \S \ref{sec:subhalo_pop}.

Moreover, the clustering of haloes have an impact on galaxy clustering. Gao, Springel \& White (2005) showed that the clustering of haloes also depends on the halo formation time, the first indication of assembly bias. Croton, Gao \& White (2007) studied the effects of assembly bias in galaxy clustering and showed that for fixed halo mass red galaxies are more clustered than blue galaxies. Other authors have studied the effects of assembly bias in galaxy clustering using other environmental dependencies of haloes such as density and studied these effects in observations (Abbas \& Sheth 2006, Skibba \etal 2006, Tinker \etal 2008a). Wang \etal (2008) studied the correlation between colour and clustering of clusters of the same mass and they found the red clusters (or with a red central galaxy) to be more clustered than blue. However, Berlind \etal (2006) seem to find the opposite result. 


On the other hand, semi-analytical models (SAM) populate galaxies in the dark matter haloes by modelling baryonic processes such as gas cooling, disk formation, star formation, supernova feedback, reionization, ram pressure or dust extinction (Cole \etal 2000, Baugh 2006, Baugh 2013) according to the potentials of dark matter. These processes contain free parameters that can be constrained by observations. Because of these processes, semi-analytical models of galaxy formation follow the evolution of the dark matter haloes, mergers, and they are more physical than HOD models in terms of environmental dependences and evolution.

In this paper we want to study the consequencies of the assumption that the galaxy population and clustering only depends on the halo mass in SAMs. We use the Millennium Simulation (Springel \etal 2005) to measure the halo bias. We also compare different definitions of halo mass. We study the public SAMs of galaxies of the Millennium Simulation to see if we can reproduce the galaxy bias from this HOD assumption, thereby assuming that the clustering of galaxies only depends on the mass of the host halo. We do this by measuring both the halo bias and the HOD in the same simulation and for the same galaxies that we want to reproduce the bias. This analysis is similar to that of Sheth \etal (2001b) but with some important differences. First of all, Sheth \etal (2001b) model the halo bias, while we use the measurement in the simulation, and we also compare it to modelling the bias. Moreover, we include the errors of the reconstructions of galaxy clustering in order to see numerically their success. Another difference with the study of Sheth \etal (2001b) is that we study several and newer SAMs in order to compare the results between them. Finally, the Millennium Simulation presents a better resolution than GIF simulations and therefore we can study smaller masses, and larger volume so we can study the $2$-halo term properly. We also focus on large scales, where no assumptions are needed for the distribution of galaxies inside the haloes. Sheth \etal (2001b) assumed the galaxies to be tracers of dark matter particles of the haloes, an assumption that has an impact on the 1-halo term of the two-point clustering.  Here we only look at the $2$-halo term. We also analyse the clustering dependence of the halo occupation of subhaloes, and its dependence on halo mass.

The paper is organized as follows. In \S \ref{sec:simulation} we introduce the Millennium Simulation as well as the semi-analytical models of galaxy formation. In \S \ref{sec:results} we present the measurements of galaxy and halo bias and the HODs that we use to reconstruct the galaxy bias and compare it to the measurements in the simulation, which is developed in \S \ref{sec:reconstructions}. We finish with a summary and discussion in \S \ref{sec:conclusions}.

\section{Simulation data}\label{sec:simulation}

For our study we use the Millennium Simulation (Springel et al.\ 2005), carried out by the Virgo Consortium using the GADGET2 (Springel \etal 2001) code with the TREE-PM (Xu 1995) algorithm to compute the gravitational interaction. The simulation corresponds to a $\Lambda CDM$ cosmology with the parameters: $\Omega_m = \Omega_{dm} + \Omega_b = 0.25$, $\Omega_b = 0.045$, $h = 0.73$, $\Omega_{\Lambda} = 0.75$, $n = 1$ and $\sigma_8 = 0.9$. It contains $2160^3 = 10,077,696,000$ particles of mass $8.6 \times 10^8\Mo$ in a comoving box of size $500\Mpc$, with a spatial resolution of $5\kpc$. The Boltzmann code CMBFAST (Seljak \& Zaldarriaga 1996) has been used to compute the initial conditions based on WMAP (Spergel \etal 2003) and 2 degree Field Galaxy Redshift Survey (2dFGRS) data (Cole \etal 2005). The simulation output starts at $z = 127$ and it has $64$ snapshots from this time to $z = 0$. 

\subsection{Haloes}

\begin{figure}
\begin{centering}
\includegraphics[width=84mm]{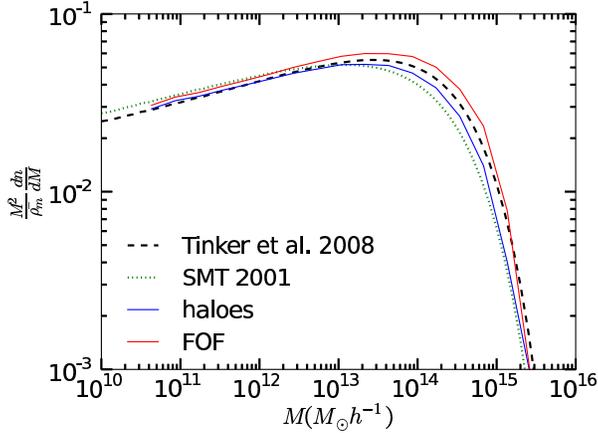}\caption[Halo and FOF Mass Function]{Normalized Mass Function of FOFs (red) and haloes (blue) compared to theoretical models. Black dashed line represent the theoretical mass function from Tinker \etal 2008. Green dotted line shows the model from Sheth, Mo \& Tormen 2001.} \label{fig:halo_mf}
\par\end{centering}
\end{figure}

In each snapshot, the haloes are identified as Friends-of-Friends (FOF) groups with a linking length of $0.2$ times the mean particle separation. All the FOF with fewer than $20$ particles are discarded. Then, subhaloes are identified in the FOF groups using the SUBFIND (Springel \etal 2001) algorithm, discarding all the subhaloes with fewer than $20$ particles. The largest object found by SUBFIND in the FOF is located in its centre, and usually has approximately the $90\%$ of the FOF mass. In SUBFIND all the particles gravitationally unbound to the subhalo are discarted, giving a gravitationally bound object. For this reason, the largest SUBFIND object can be seen as the gravitationally bound core of the FOF, and in this paper we will call it halo.
We must notice that these FOF and halo catalogues are independent of the galagxy catalogues of the simulation. In Fig. \ref{fig:halo_mf} we show the mass function of FOF and haloes, in red and blue lines respectively, compared to the theoretical models of Tinker \etal (2008b) and Sheth, Mo \& Tormen (2001) (hereafter SMT 2001). The mass function has been multiplied by $M^2/\bar{\rho}$ for clarity. We define the masses of the FOF as the total number of particles belonging to the FOF, and the halo mass as the total number of particles belonging to the halo (the largest SUBFIND of the FOF). We can see that the haloes have a mass function close to the model of SMT 2001, where the ellipsoidal collapse model has been used. On the other hand, the mass function of FOF is closer to the model of Tinker \etal (2008). We can also see that the mass function of haloes is very close to the one of FOF at low masses, meaning that most of the small  haloes are contained in small FOFs, while at large masses the mass function of haloes is lower, meaning that the differences in the mass definition of haloes and FOFs become larger.

\subsection{Galaxies}

Galaxy catalogues of several semi-analytical models (SAM) are available in the public database of the simulation. In this paper we study several SAMs (Bertone, De Lucia \& Thomas 2007, Bower \etal 2006, De Lucia \& Blaizot 2007, Guo \etal 2011, Font \etal 2008). In the models of Bertone, De Lucia \& Thomas 2007, De Lucia \& Blaizot 2007 and Guo \etal 2011 (BDLT07, DLB07 and G11 respectively hereafter), developed at the Max-Planck-Institute for Astrophysics (MPA) in Garching\footnote{\url{http://gavo.mpa-garching.mpg.de/MyMillennium/}}, the galaxies are placed and evolved in the subhaloes according to their properties and their merger trees (Springel \etal 2005, Croton \etal 2006, De Lucia \& Blaizot 2007). In the case of Bower \etal 2006 and Font \etal 2008 models (B06 and F08 hereafter), developed at the Institute for Computational Cosmology in Durham\footnote{\url{http://galaxy-catalogue.dur.ac.uk:8080/MyMillennium/}}, the merger trees are constructed using the \textit{Dhaloes}, a different definition of halo consisting in groups of subhaloes (Helly \etal 2003, Jiang \etal 2013). In most of the cases, the Dhaloes consist in the set of subhaloes of the same FOF, but in some cases \footnote{if the subhalo is outside twice the half mass radius of the parent halo or the subhalo has retained $75\%$ of the mass it had at the last output time where it was an independent halo.} these sets are divided in different Dhaloes (see Merson \etal 2013, Harker \etal 2006). Then the evolution of the latter models is related to halo (Dhalo) evolution, while the first models are associated to subhaloes. 

\begin{figure}
\includegraphics[width=84mm]{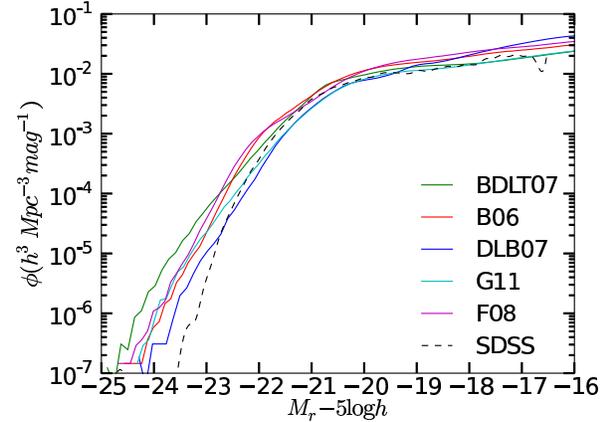}
\caption[Luminosity function of SAMs compared to SDSS]{ Luminosity Function of SAMs compared to the SDSS DR2 data (Blanton \etal 2005). Solid lines represent the different SAMs, and the dashed line shows the Luminosity Function from SDSS (Blanton \etal 2005). } 
\label{fig:lf_vs_sdss}
\end{figure}

In Fig.\ \ref{fig:lf_vs_sdss} we compare the luminosity function of all the SAMs studied with the luminosity function of SDSS DR2 (Blanton \etal 2005). To make coherent comparisons, we have used the luminosities in SDSS $r$ filter of galaxies including dust extinction of the SAMs. We applied a factor of $5 \log h$, with $h = 0.73$, in the MPA models, since this factor was not included in the database. In this Figure we can see an evident excess of bright galaxies in all the models. There is also a slight excess of galaxies in the faint end in the models DLB07, B06 and F08. However, Bernardi \etal (2013) studied different algorithms to obtain galaxy magnitudes and argued that the luminosity function from Blanton \etal (2003) in $r$ band is probably underestimated for galaxies brighter than $M_r < -22$, so these differences do not necessarily reflect problems for the SAMs. These results are in agreement with Contreras \etal 2013, where they present a deeper comparison of the different semi-analytical models of the Millennium Simulation. Contreras \etal (2013) studied how much the clustering and HOD of these semi-analytical models depend on stellar mass, cold gas mass and star formation rate.

\section{Bias and HOD}\label{sec:results}

In this section we present our measurements of FOF, halo and galaxy bias that we will use to analyse if galaxy bias depends only on halo mass. We estimate the two-Point Correlation Function (2PCF) using density pixels and the expression
\begin{equation}
\xi(r_{12}) = \langle \delta(r_1) \delta(r_2) \rangle ,
\end{equation}\label{eq:xi}
where $\delta(r)$ refers to the density fluctuation defined by $\delta(r) = \rho(r) / \bar{\rho} - 1$ in pixels. From that, we measure the bias using the local bias model (Fry \& Gazta\~{n}aga 1993): 
\begin{equation}
b(r) = \sqrt{\frac{\xi_g(r)}{\xi_m(r)}}
\end{equation}
where $\xi_g(r)$ corresponds to the 2PCF of the studied object (haloes or galaxies), $b(r)$ is the bias factor, and $\xi_m(r)$ is the 2PCF of the dark matter field. As we assume $b(r)$ to be constant at large scales, where $\delta \ll 1$, we define the mean value by fitting $b(r)$ to a constant in the scale range $20\Mpc < r < 30\Mpc$. Although theoretically the bias may not be in the linear regime for these scales, we have checked that it behaves as a constant, so our fit can be assumed as valid. Moreover, the size of the Millennium Simulation does not allow us to go to larger distances with precision. The errors are measured with a Jack-Knife method (Norberg \etal 2009) of this measurement of $b$, using $64$ cubic subsamples. The errors are taken from the standard deviation of these subsamples. The distribution of these subsamples is close to a gaussian, and the errors obtained from the percentiles are very similar to those from the standard deviation.

\subsection{Bias}\label{sec:bias}

\begin{figure}
\begin{centering}
\includegraphics[width=84mm]{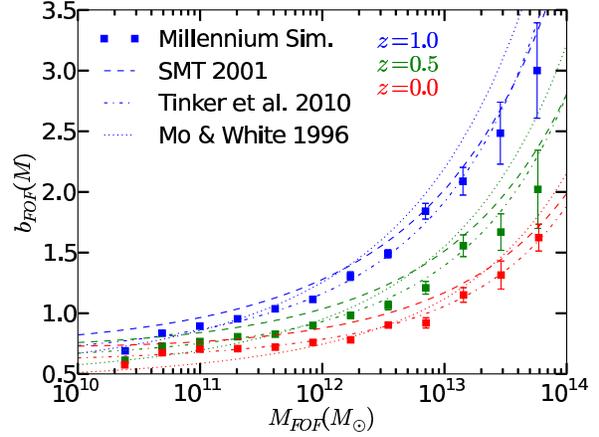}
\includegraphics[width=84mm]{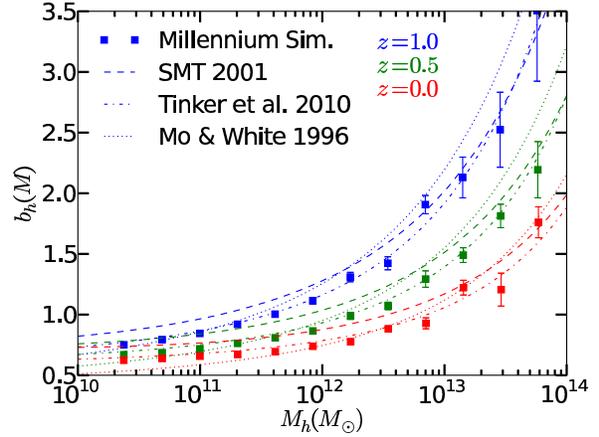}\caption[FOF and halo bias as a function of mass]{FOF (top) and halo (bottom) bias as a function of mass at $3$ different redshifts compared to theoretical expressions. The squares show the measurements of bias from the Millennium Simulation. Dashed lines show the analytic model from Sheth, Mo and Tormen (2001), dashed-dotted lines correspond to the Tinker \etal (2010) model and dotted lines are the analytic expressions from Mo \& White (1996). Each colour represents a different redshift, as specified.} \label{fig:hal_bias}
\par\end{centering}
\end{figure}

In Fig.\ \ref{fig:hal_bias} we present the FOF (top) and halo (bottom) bias as a function of mass. We refer to them as $b_{FOF}(M)$, $b_h(M)$ and we compare the results with some analytical models. Using the mass function developed by Press \& Schechter (1974) assuming the spherical collapse model, Mo \& White (1996) derived the following expression for the halo bias:
\begin{equation}
b(\nu) = 1 + \frac{\nu^2 - 1}{\delta_c} ,
\label{eq:smt01}
\end{equation}
where $\delta_c = 1.686$ is the linear density of collapse and $\nu = \delta_c/\sigma(M)$, where $\sigma(M)$ is the linear rms mass fluctuation in spheres of radius $r = (3M/4\pi \bar{\rho})^{1/3}$. Sheth, Mo \& Tormen (2001) (SMT 2001) generalized and improved the expression using an ellipsoidal collapse model and they obtained the result: 
\begin{equation}
\begin{split}
b(\nu) = 1 + \frac{1}{\sqrt{a} \delta_c}  \sqrt{a} (a \nu^2) + \sqrt{a} b (a \nu^2)^{1-c} \\  
- \frac{(a \nu^2)^c}{(a \nu^2)^c + b(1-c)(1-c/2)}  ,
\end{split}
\label{eq:mw96}
\end{equation}
with the parameters $a = 0.707$, $b = 0.5$ and $c = 0.6$ tuned to work in N-body simulations. Finally, Tinker \etal (2010) presented a more flexible expression: 
\begin{equation}
b(\nu) = 1 - A \frac{\nu^a}{\nu ^a + \delta_c^a} + B \nu^b + C \nu^c .
\label{eq:tinker2010}
\end{equation}
The values of the parameters of this expression used in our comparisons correspond to the values shown in Table $2$ of Tinker \etal (2010) with $\Delta = 200$.

First of all, we can see that the SMT 2001 model tends to overpredict $b_{FOF}(M)$ and $b_h(M)$, especially at low masses. We can see a difference in the high mass region between FOF and haloes because for each FOF the mass of the halo is reduced (and then shifted to a lower mass) due to the SUBFIND unbinding process. On the other hand, Mo \& White (1996) model tends to produce an overprediction at high masses and an underprediction at low masses for all the cases. One possible reason for this is that Mo \& White (1996) assume the Press-Schechter mass function, but this mass function fails to reproduce the halo mass function in simulations (Gross \etal 1998, Governato \etal 1999, Lee \& Shandarin 1999, Sheth \& Tormen 1999, Jenkins \etal 2001). Finally, the agreement of the Tinker \etal (2010) expression with $b_{FOF}(M)$ and $b_h(M)$ is remarkable.

\begin{figure*}
\includegraphics[width=84mm]{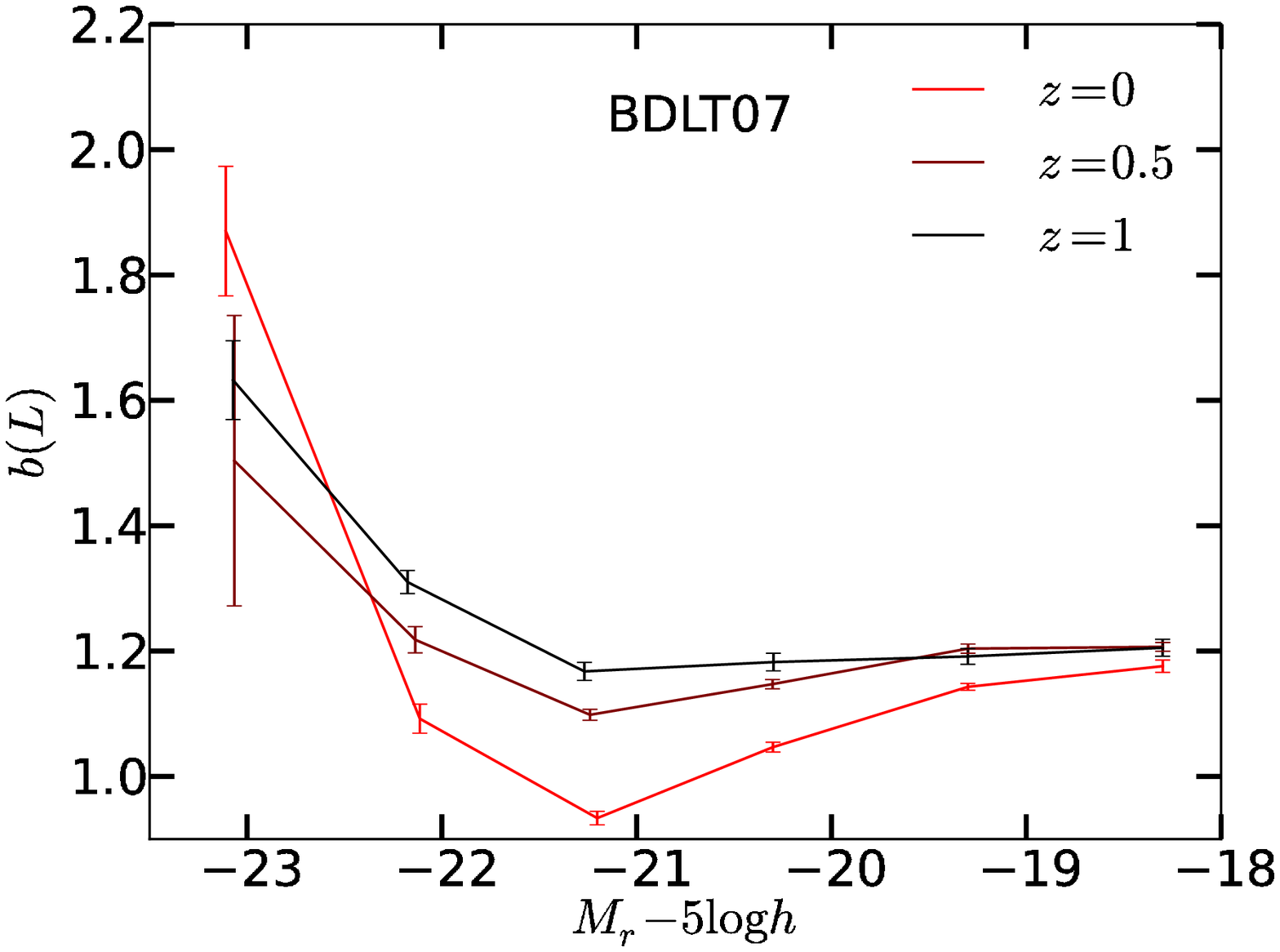}
\includegraphics[width=84mm]{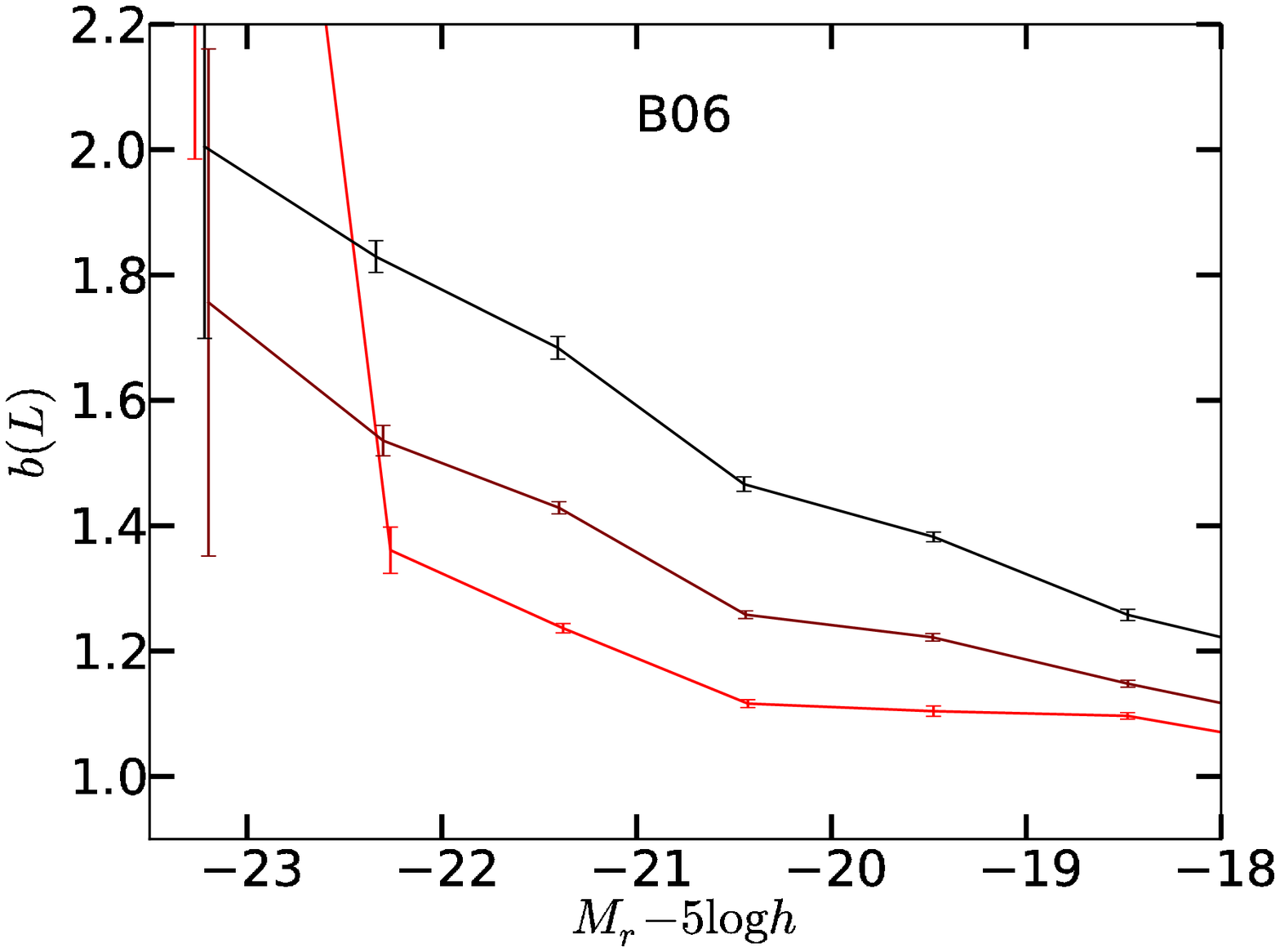} 
\includegraphics[width=84mm]{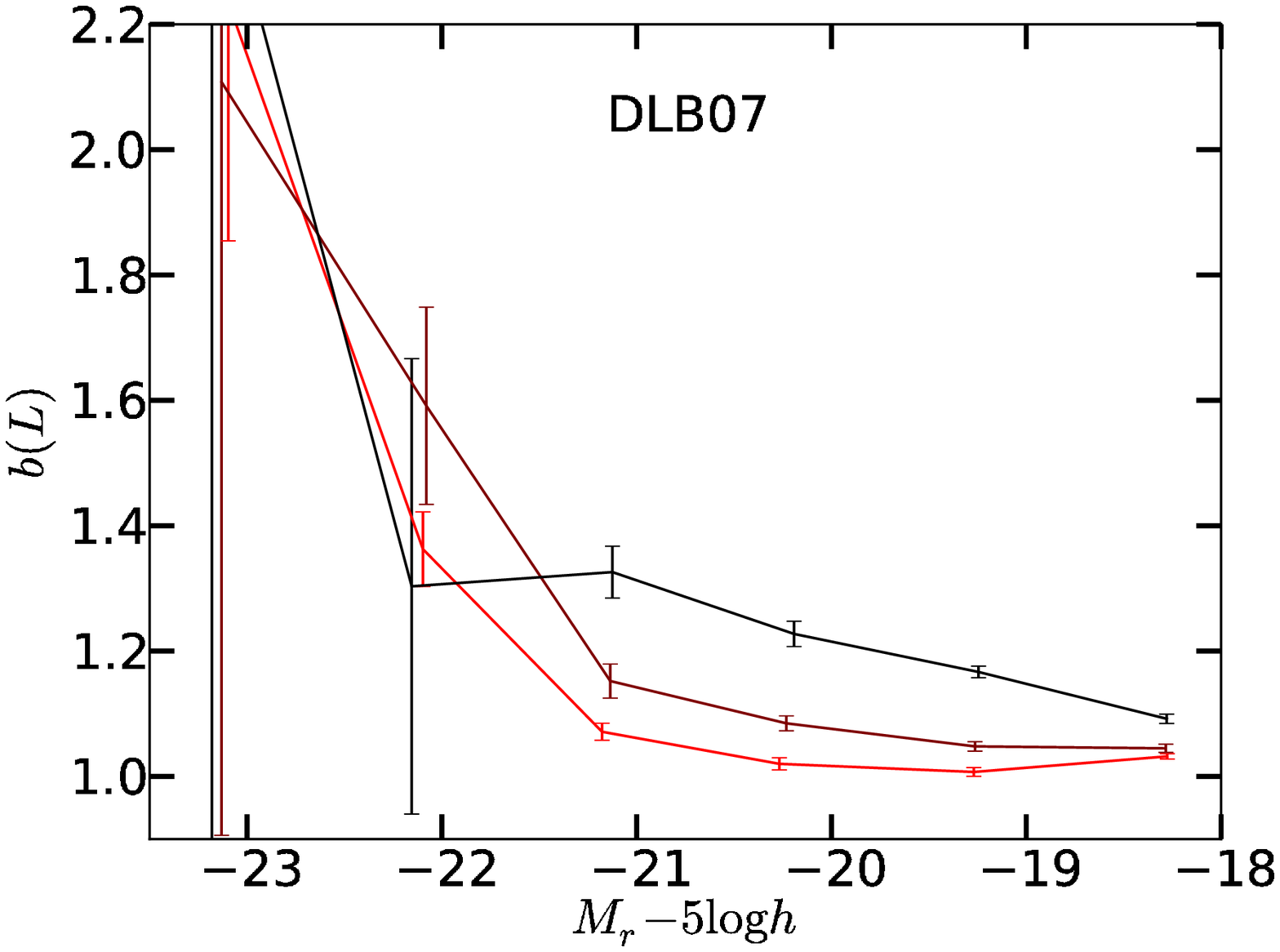}
\includegraphics[width=84mm]{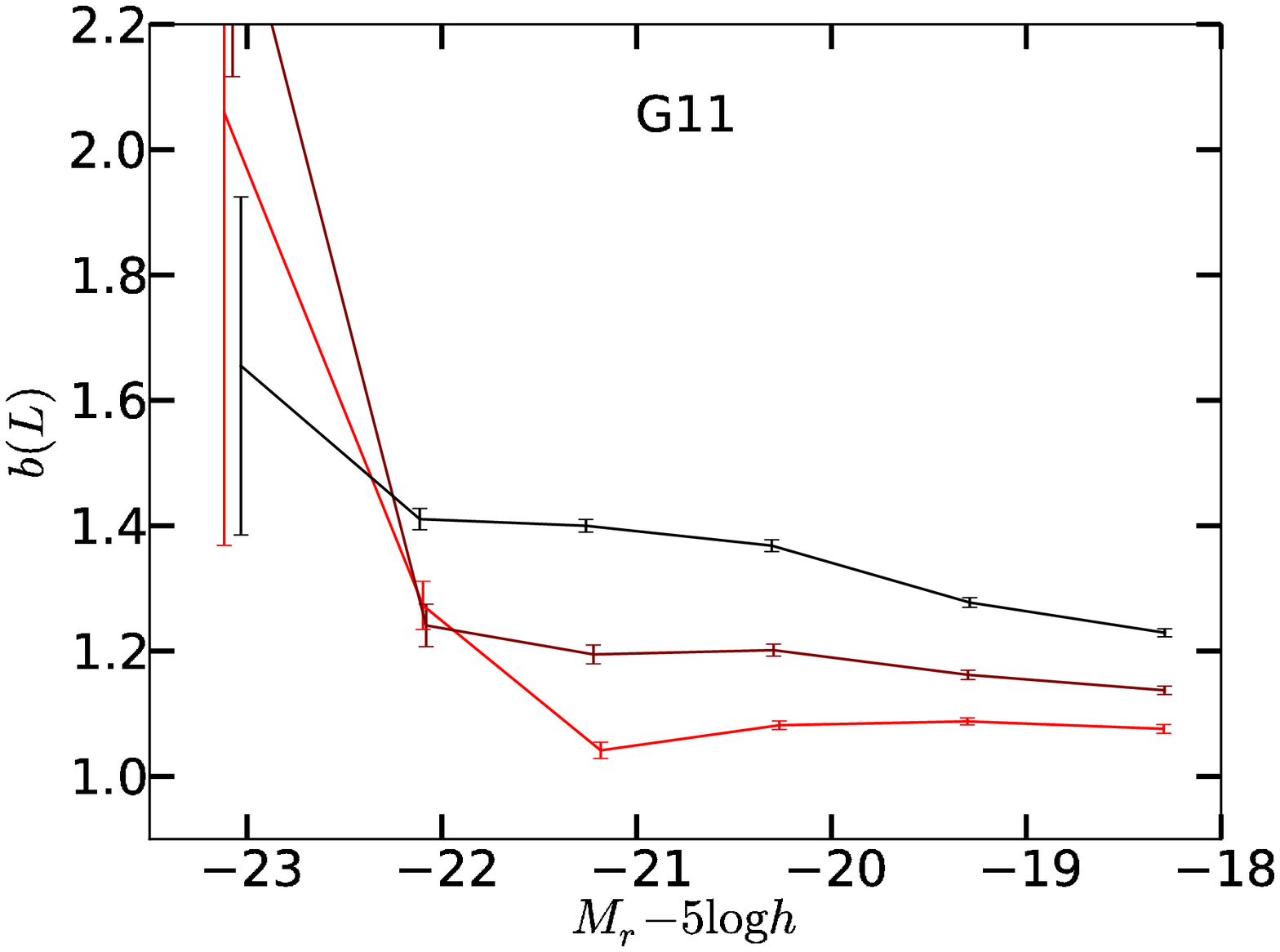}
\includegraphics[width=84mm]{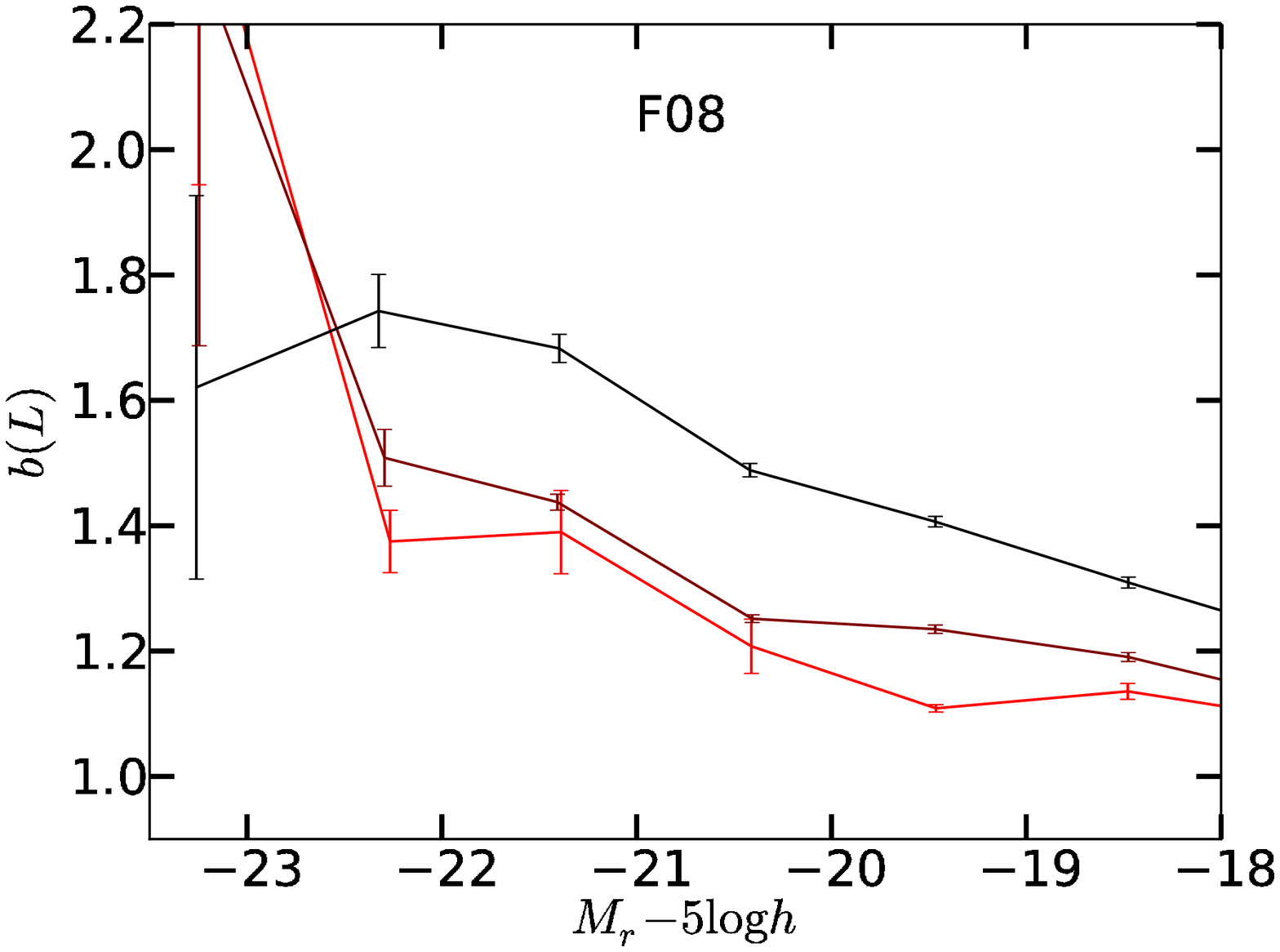}
\includegraphics[width=84mm]{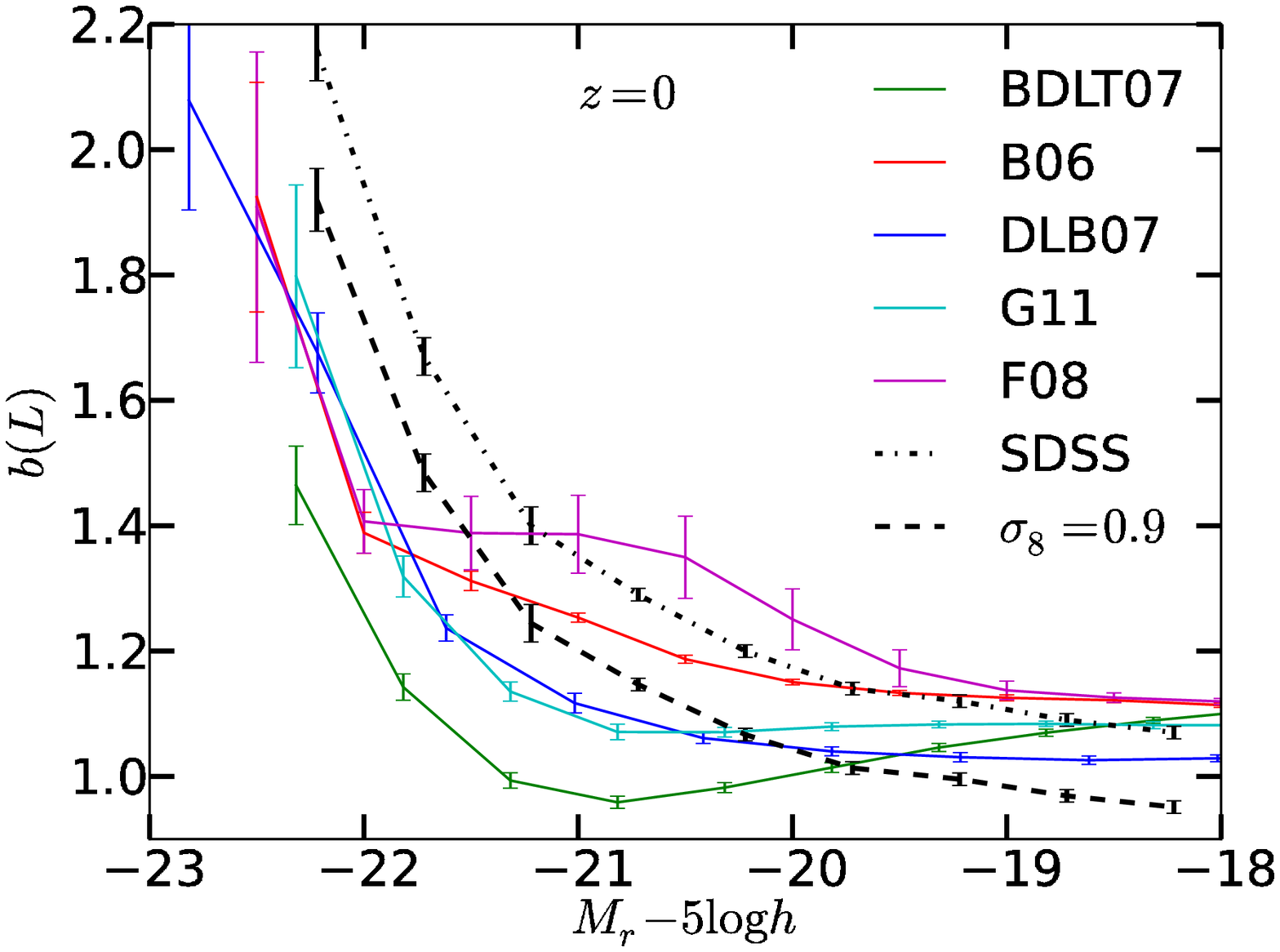}
\caption[Galaxy bias of SAM at different $z$]{Luminosity dependence (in absolute $r$ band magnitude) of galaxy bias for $5$ SAMs and comparison with observations (bottom-right panel). The first $5$ panels correspond to $b_g(L)$ at $z = 0, 0.5, 1$ for each SAM as labelled in magnitude bins. In the bottom right panel, $b_g(L)$ of all the SAMs at $z = 0$ are compared to $b_g(L)$ from SDSS DR7 (Zehavi \etal 2011) in magnitude thresholds instead of bins. Solid lines represent the Millennium catalogues. The black dashed-dotted line corresponds to the SDSS DR7 data. The black dashed line shows a correction of $0.8/0.9$ to approximate the amplitude of $b_g(L)$ from SDSS if the 2PCF were normalized to a cosmology of $\sigma_8 = 0.9$ instead of $\sigma_8 = 0.8$.} 
\label{fig:gal_gen_bias}
\end{figure*}

In Fig.\ \ref{fig:gal_gen_bias} we show galaxy bias as a function of luminosity, $b_g(L)$, for the SAMs. The first $5$ panels show $b_g(L)$ in bins of luminosity for each of the studied SAMs, in the $r$ band at $3$ different redshifts, as specified. We also show in bottom right panel the comparison of these models with observations in the SDSS DR7 (Zehavi \etal 2011) using luminosity thresholds.  To compute $b_g(L)$ from the SDSS DR7 data, Zehavi \etal (2011) used a prediction of the dark matter correlation function from a $\Lambda CDM$ cosmological model (Smith \etal 2003). Solid lines show the different SAMs, while the dashed-dotted line shows $b_g(L)$ from Zehavi \etal (2011). As the cosmology assumed in Zehavi \etal (2011) is different than the one from Millennium, the dashed line corresponds to multiply the SDSS measurement by a factor $0.8 / 0.9$. This factor is an approximation of the difference in the amplitude of the dark matter field of both cosmologies if we assume that $b_g(L)$ behaves as $\sigma_8$. Then, the dashed line shows an approximation of $b_g(L)$ of the SDSS galaxies normalized by the Millennium cosmology. 

First of all, we can see that $b_g(L)$ increases with $z$, although the brightest galaxies tend to show higher clustering amplitude at low $z$. From the last panel, however, we observe discrepancies for all the models with observations from SDSS DR7 presented in Zehavi \etal (2011). We notice the good agreement between F08 and B06 models in the brightest galaxies. In general, the predicted $b_g(L)$ is lower than in the observations for the brightest galaxies, and the shape of $b_g(L)$ steepens only for the brightest galaxies in the SAMs, showing a shift with respect to the SDSS DR7 data. This shift depends strongly on the different cosmologies adopted between the SDSS analysis and the Millennium Simulation. As the value of $\sigma_8$ is higher in the Millennium Simulation, their galaxy clustering is underpredicted due to the lower clustering of the dark matter of the simulation. The dashed line shows the comparison between the models and SDSS assuming the same dark matter field cosmology parameter $\sigma_8$. Here we can see that the agreement is better in the brightest galaxies, but worse in the faint end. From this panel, we can say that the agreement on $b_g(L)$ between the models and the observations is strongly dependent on the cosmologies that we assume, but anyway the shape of $b_g(L)$ of the models is different than that of SDSS data. If $M_r$ in SDSS is shifted up as suggested by the luminosity function of Fig.\ \ref{fig:lf_vs_sdss}, then the agreement will be better.

\subsection{HOD}\label{subsec:HOD}

\begin{figure*}
\includegraphics[width=84mm]{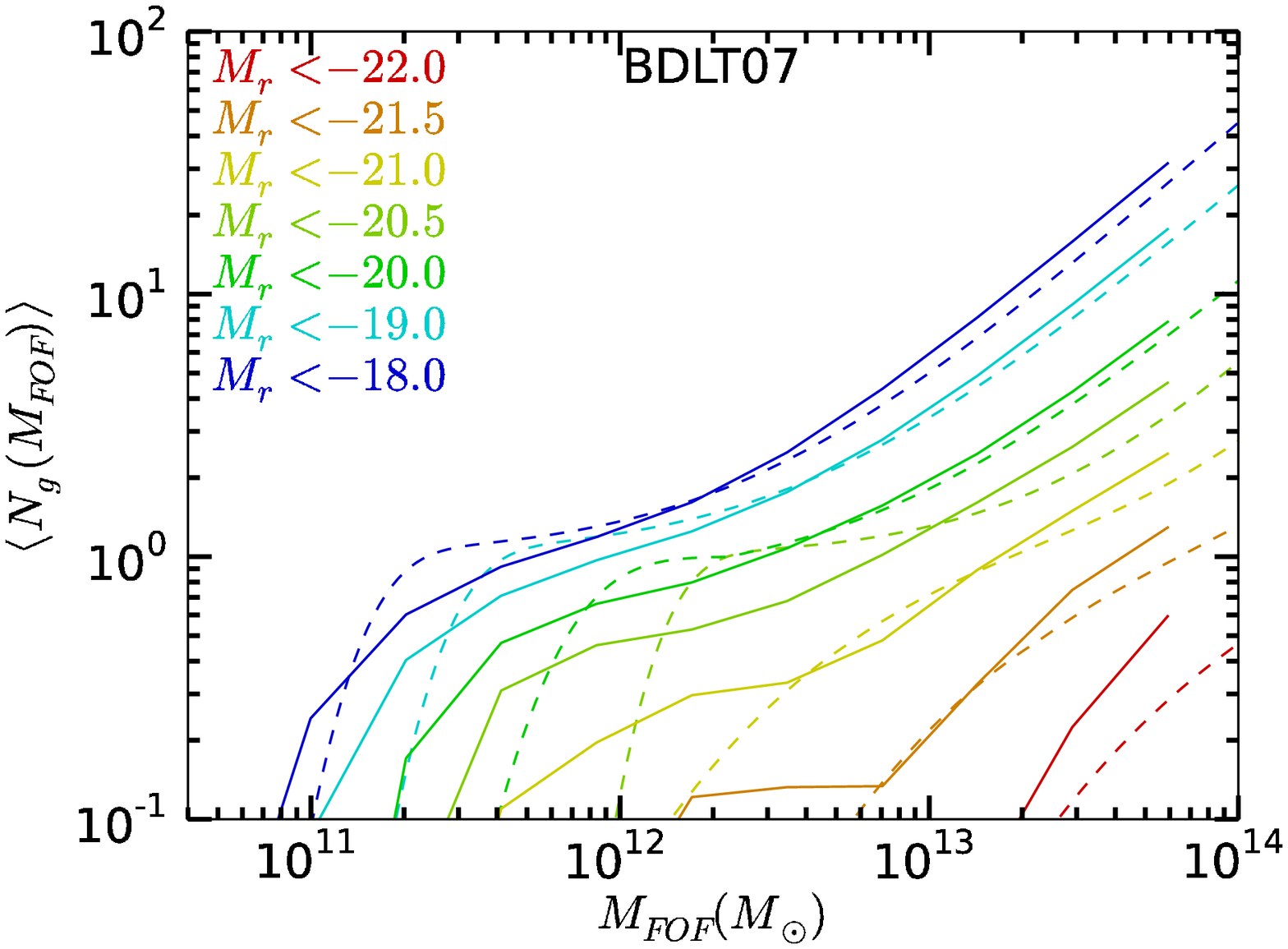}
\includegraphics[width=84mm]{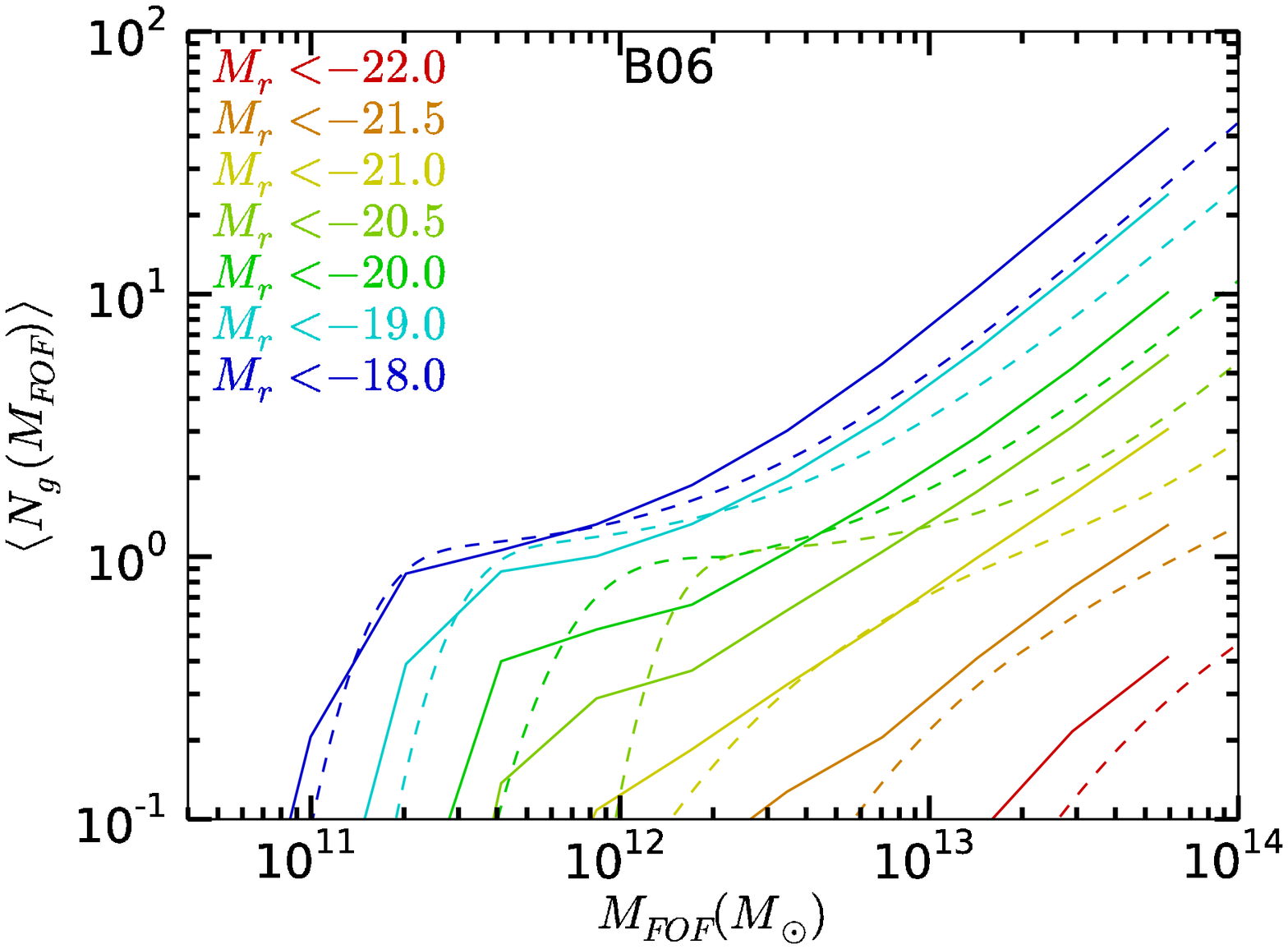}
\includegraphics[width=84mm]{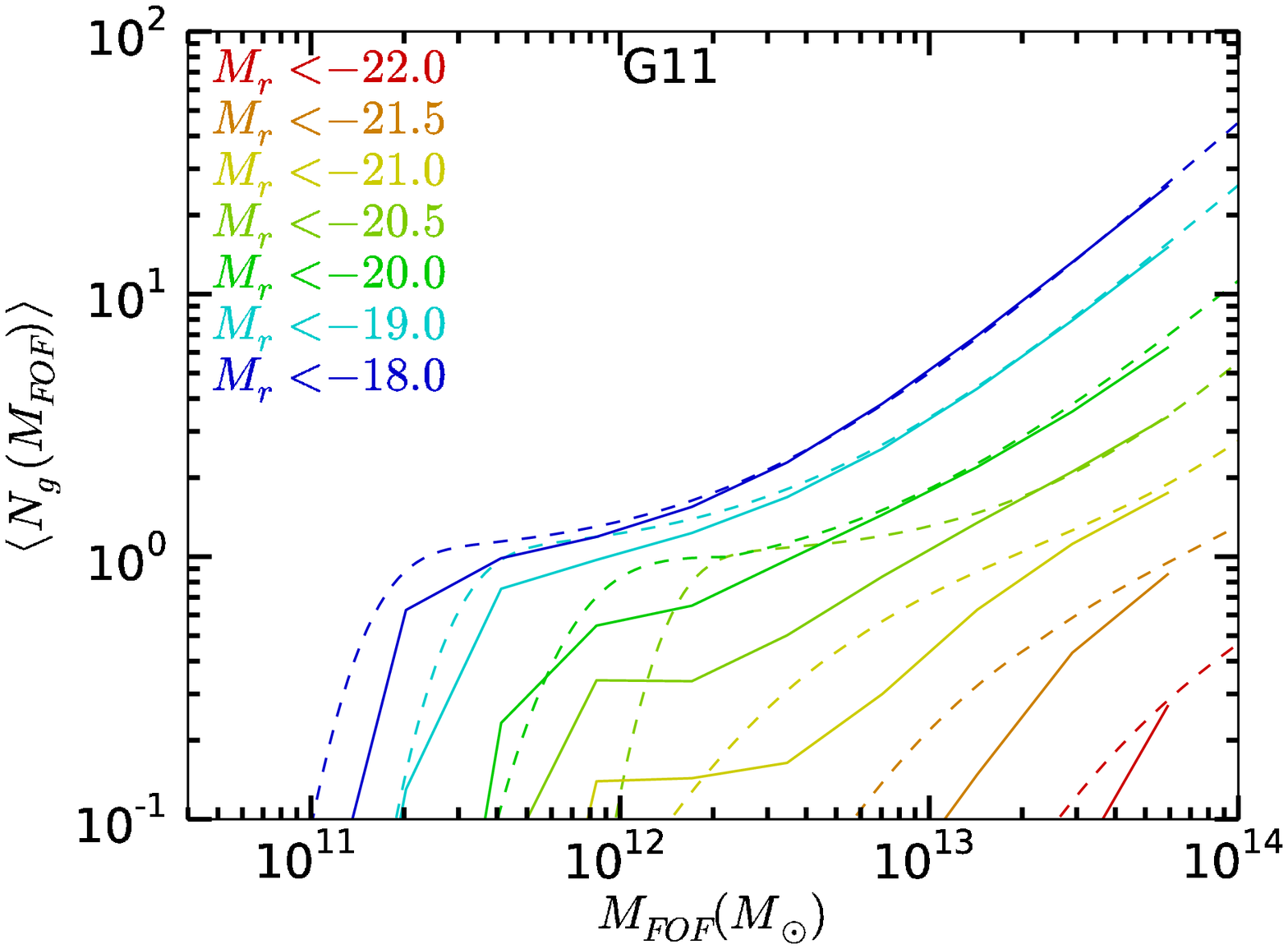}
\includegraphics[width=84mm]{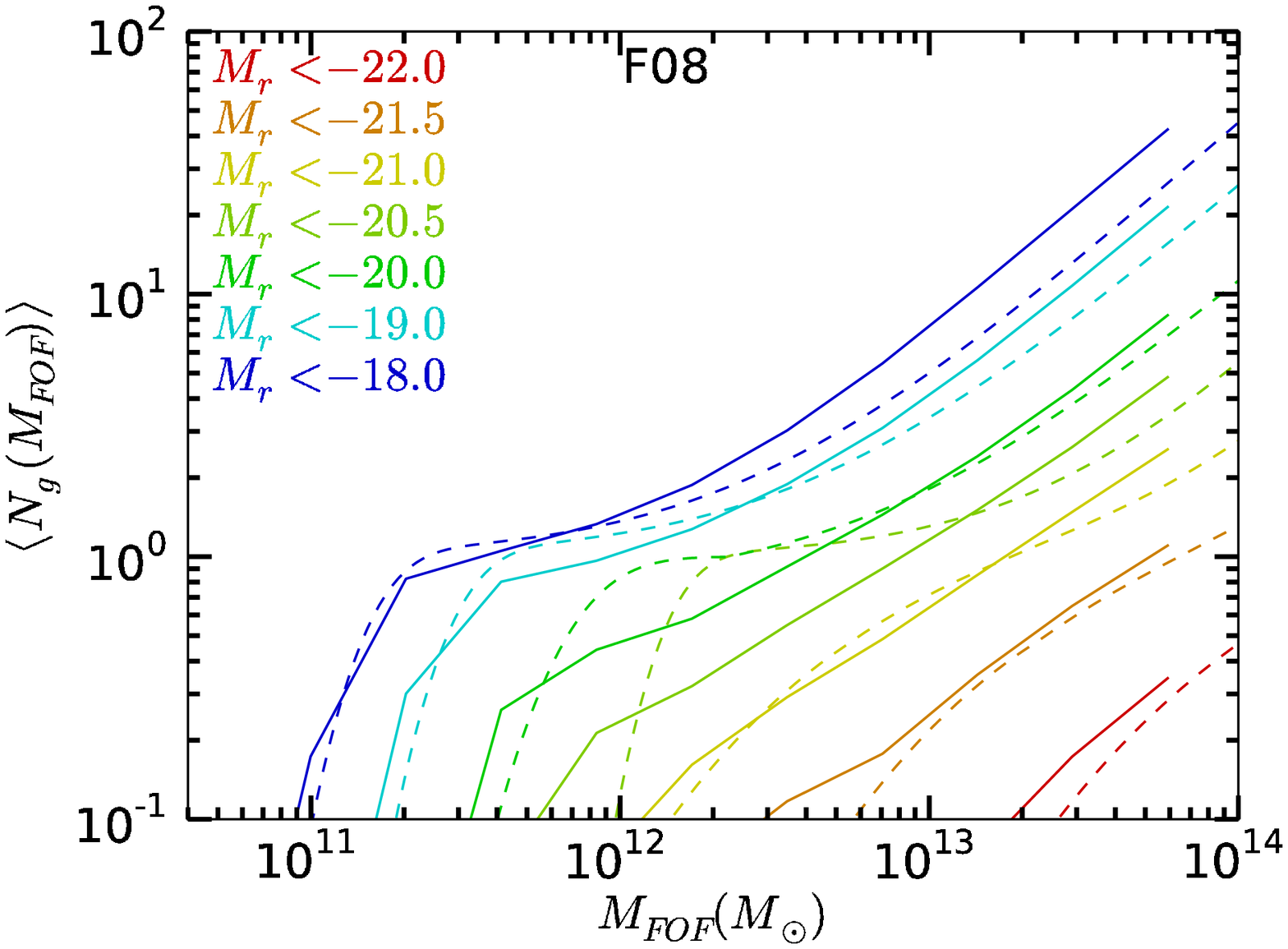}
\caption[Comparison of HOD between SAMs and SDSS]{HOD of galaxies for the different SAMs (solid lines) compared to the HOD found by Zehavi \etal (2011) from the SDSS DR7 data (dashed lines). Top panels correspond to the BDLT07 (left) and B06 (right) models, bottom panels show G11 (left) and F08 (right) models. Each colour corresponds to a luminosity threshold in units of $M_r - 5 \log h$ as specified.} 
\label{fig:hod_vs_sdss}
\end{figure*}

SAMs of galaxy formation are not based on the halo model and hence they do not use the HOD prescription to populate galaxies into haloes. But effectively the models produce an HOD as an output, and we can measure the occupation of galaxies in haloes and study the mass dependence of the different populations. For each galaxy catalogue, we calculate the HOD by counting the galaxies per halo as a function of the halo mass. For the reconstructions of $b_g(L)$ in \S \ref{sec:reconstructions}, we will assume the HOD to be only dependent of halo mass (FOF or halo mass). We also analyse the luminosity dependence of these HODs. These distributions are shown in Fig.\ \ref{fig:hod_vs_sdss}, where the HOD of some models are compared to the measurements from SDSS DR7 (Zehavi \etal 2011). Zehavi \etal (2011) inferred the HOD measurements from the clustering of different samples of galaxies assuming that the clustering of galaxies can be expressed in terms of the probability distribution that a halo of a given virial mass $M$ hosts $N$ galaxies of a given type. In this calculation they assume a $\Lambda$CDM cosmology with $\Omega_m = 0.25$, $\Omega_b = 0.045$, $\sigma_8 = 0.8$, $H_0 = 70 \, {\rm km} \, {\rm s}^{-1} \, {\rm Mpc}^{-1}$ and $n_s = 0.95$. Dashed lines in Fig.\ \ref{fig:hod_vs_sdss} show the best fit of the HODs of the SDSS RD7 presented in Zehavi \etal (2011) using the equation:

$\langle N(M_h) \rangle = $
\begin{equation}
\frac{1}{2} \left[ 1 + \mbox{erf} \left( \frac{\log M_h - \log M_0}{\sigma_{\log M}} \right) \right]\left[ 1 + \left(\frac{M_h - M_0}{M'_1}\right)^\alpha \right] ,
\end{equation}
where $M_{min}$, $\sigma_{\log M}$, $M_0$, $M'_1$ and $\alpha$ are parameters to be fitted.  The SAM measurements are shown by solid lines and each colour corresponds to a different threshold in magnitude, $M_r$, using the FOF mass.  The values of the HOD of the SAMs using the haloes instead of the FOF groups, although it is not shown, is pretty similar. Given the fact that the haloes have always lower mass than their respective FOF (since the difference with respect to the FOF is due to the application of the unbinding processes for dark matter particles), $N(M_{h})$ always has to be higher than $N(M_{FOF})$ for a given mass if the slope of the HOD is positive, as it is. At high masses, we note that although the original SAMs tend to have a higher population, the agreement with observations is remarkable, and very good in the particular case of G11 model. At low masses, the level of agreement is model dependent, but in general the change of slope at $\langle N_g (M_{FOF}) \rangle \approx 1$ tends to be softer in the SAMs than in the SDSS DR7 measurements. 

We must mention, however, that the HOD measurements depend on cosmology, and different assumptions on cosmology can give different fits of the HOD in observations. Since the cosmology assumed in Zehavi \etal 2011 is different than the Millennium cosmology, the HOD measurements of both cases do not need to agree.

\section{Galaxy bias reconstructions}\label{sec:reconstructions}

\begin{figure*}
\begin{centering}
\includegraphics[width=84mm]{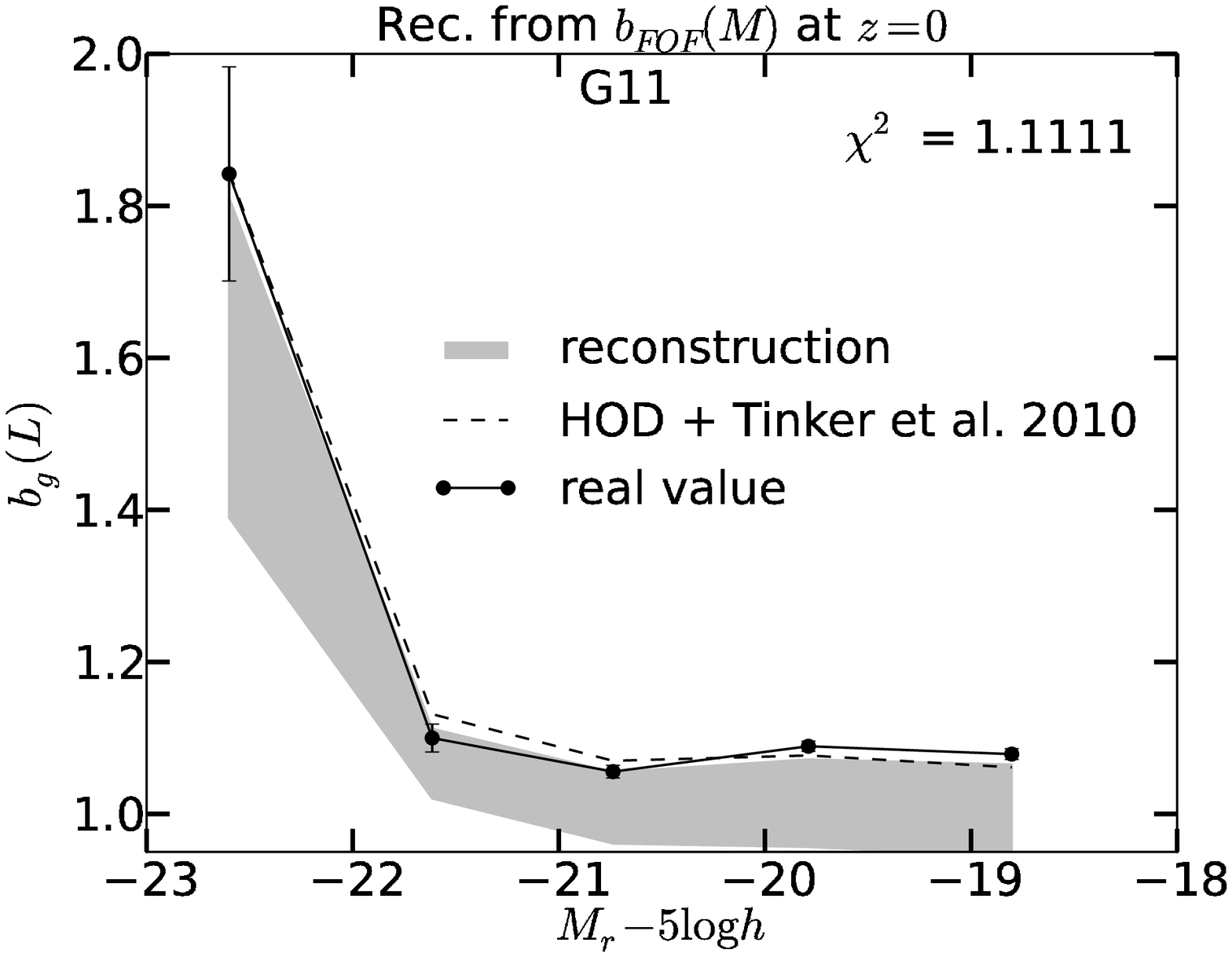}
\includegraphics[width=84mm]{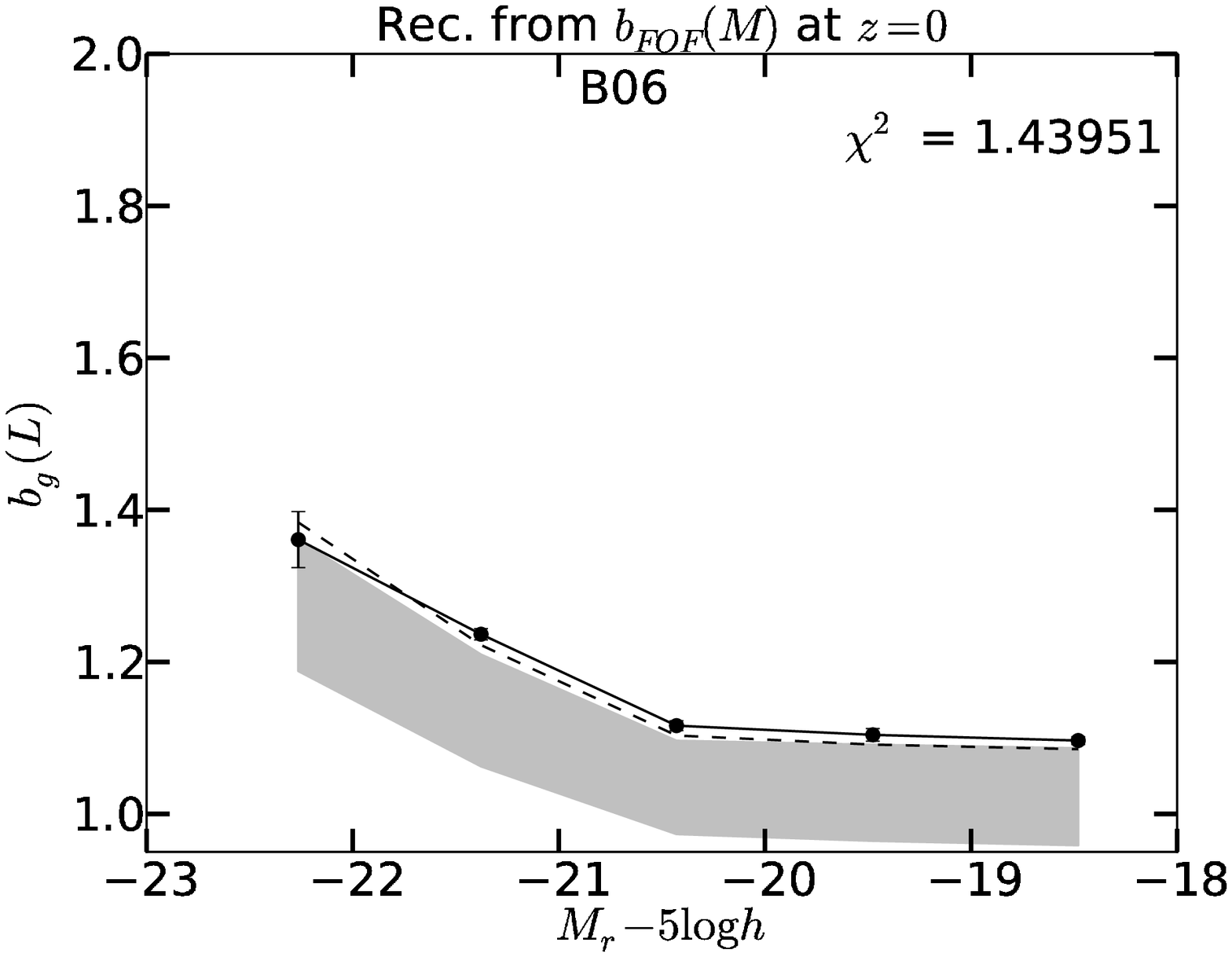}
\includegraphics[width=84mm]{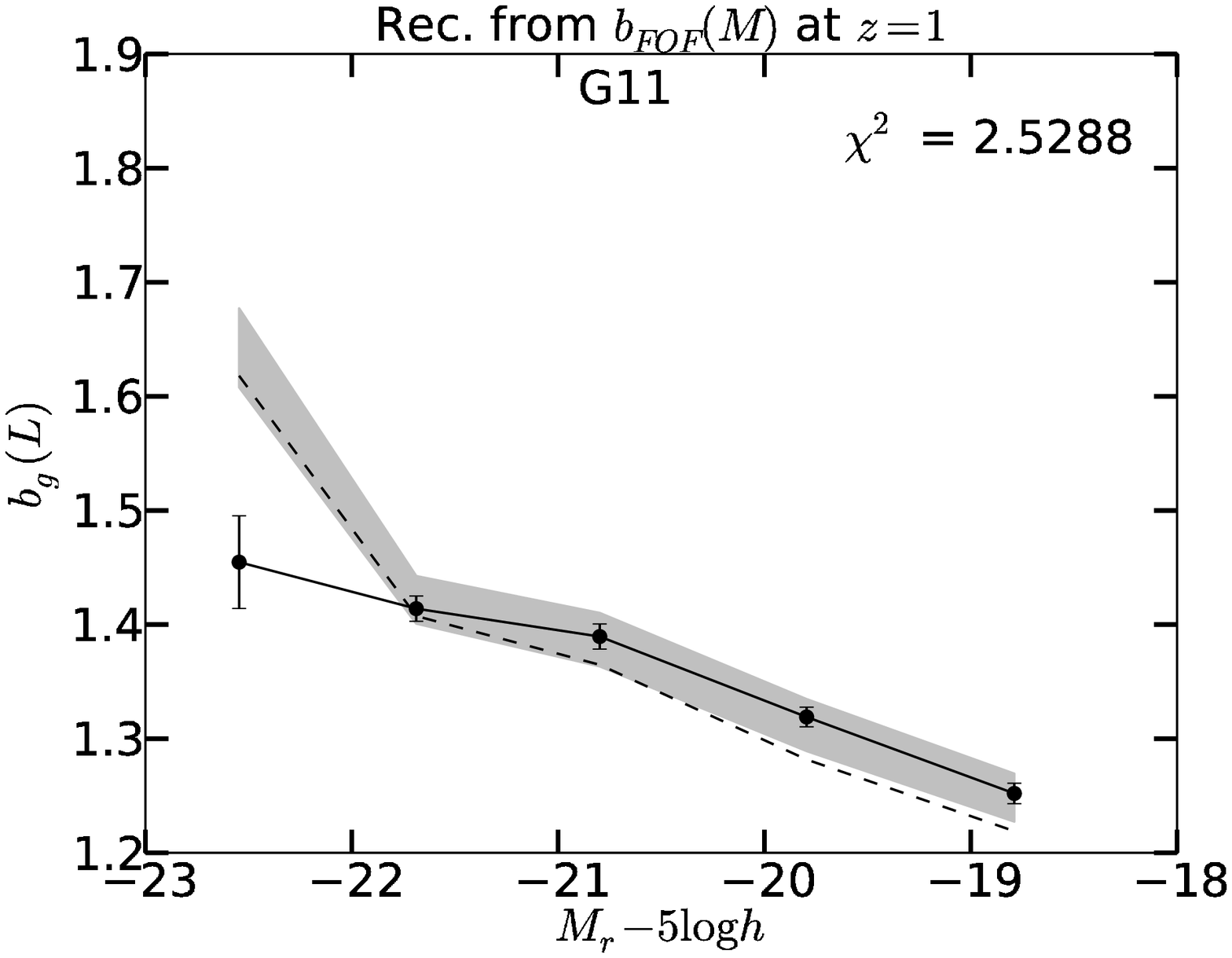}
\includegraphics[width=84mm]{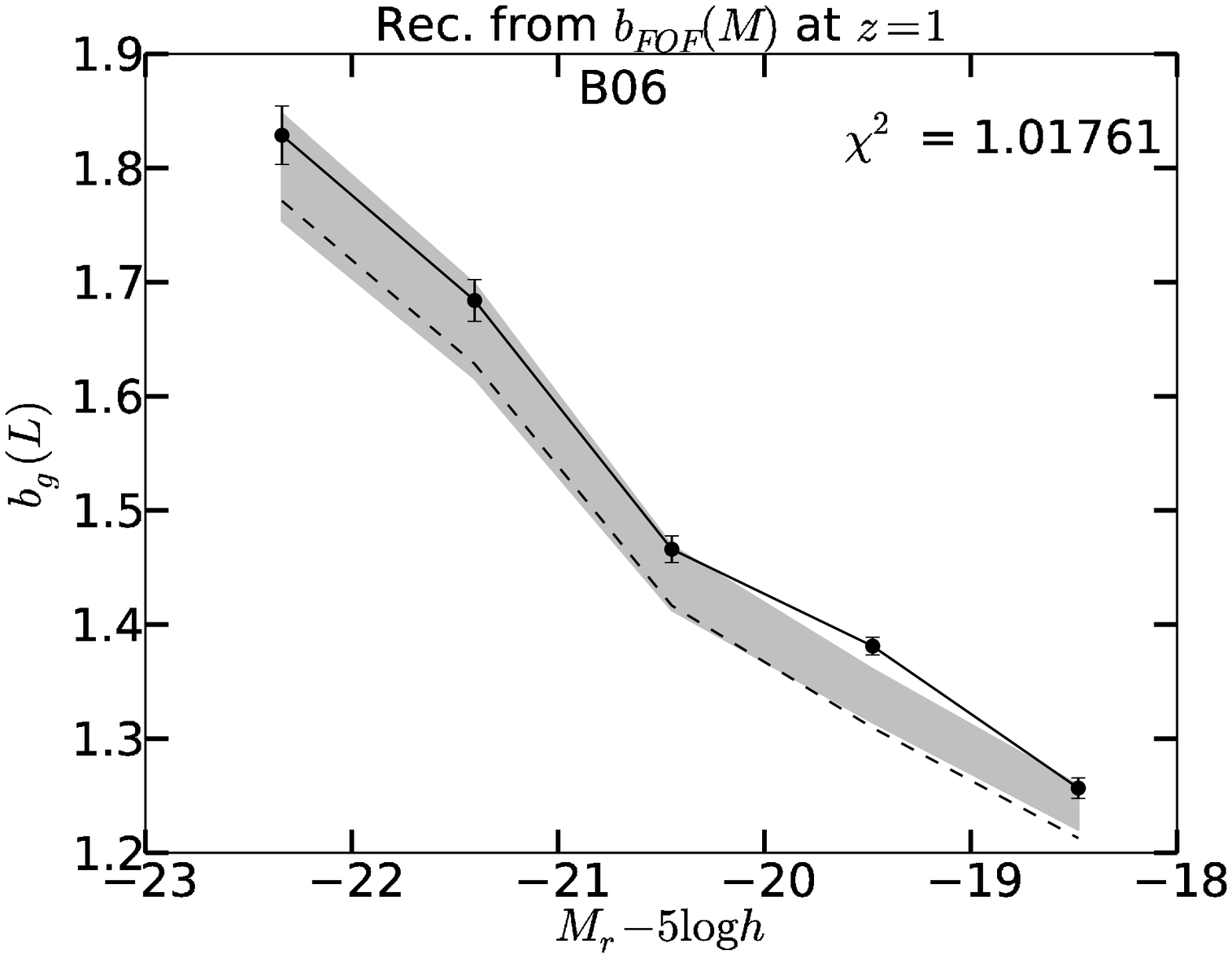}
\caption[Reconstruction of galaxy bias from halo bias]{Reconstructions of $b_{g}(L)$ from $b_{FOF}(M)$ at $z = 0$ and $z = 1$. The grey shaded region corresponds to the predicted $b_{rec}(L) \pm 1 \sigma$, while solid line represents the real measured value of $b_{g}(L)$, and the dashed line corresponds to the reconstructions using the values of $b_{FOF}(M)$ from the Tinker \etal (2010) model. Left panels corresponds to G11 model, while right panels show predictions for the B06. On top, $z = 0$. On bottom, $z = 1$.}
\label{fig:rec_gal_fof}
\par\end{centering}
\end{figure*}

\begin{figure*}
\begin{centering}
\includegraphics[width=84mm]{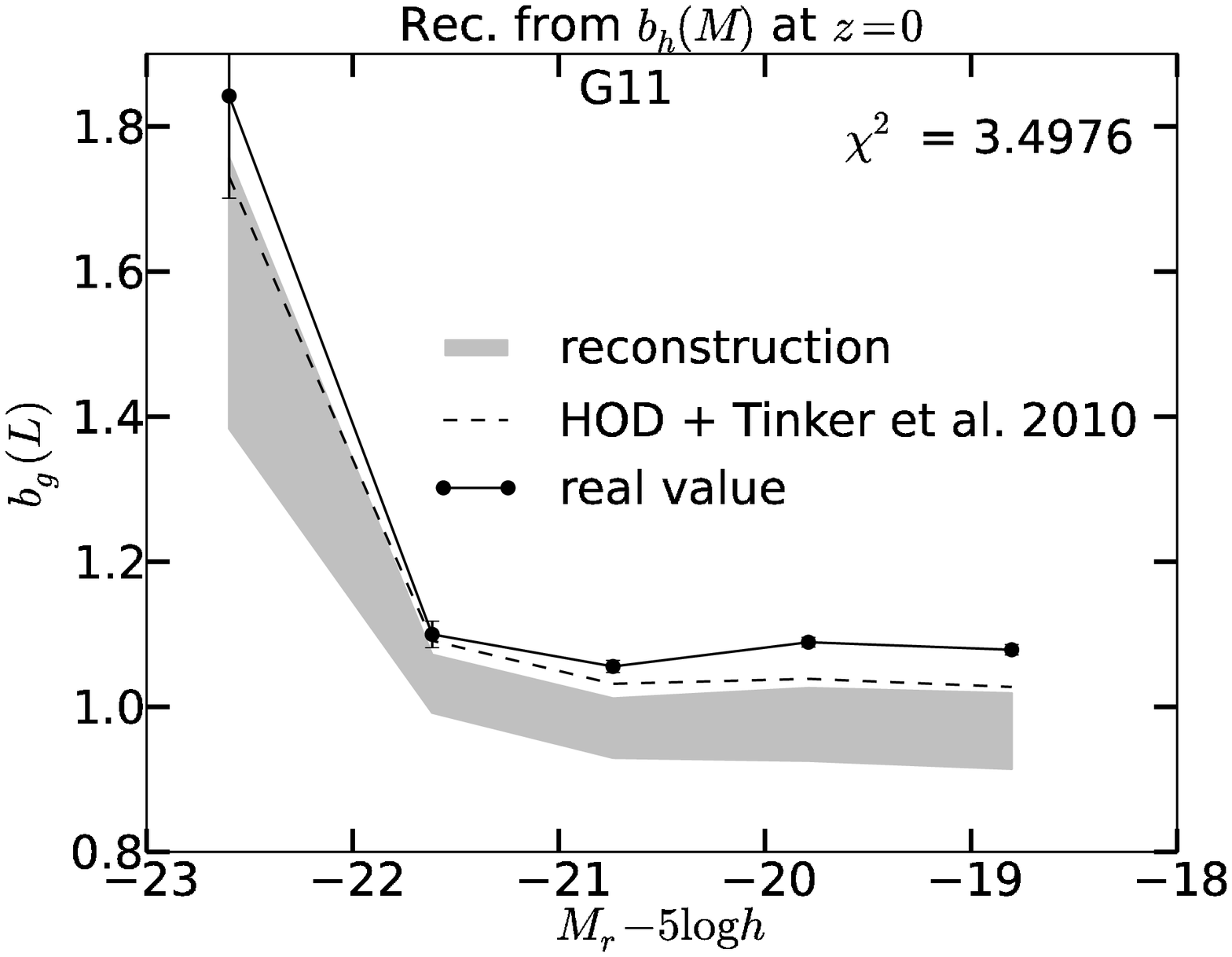}
\includegraphics[width=84mm]{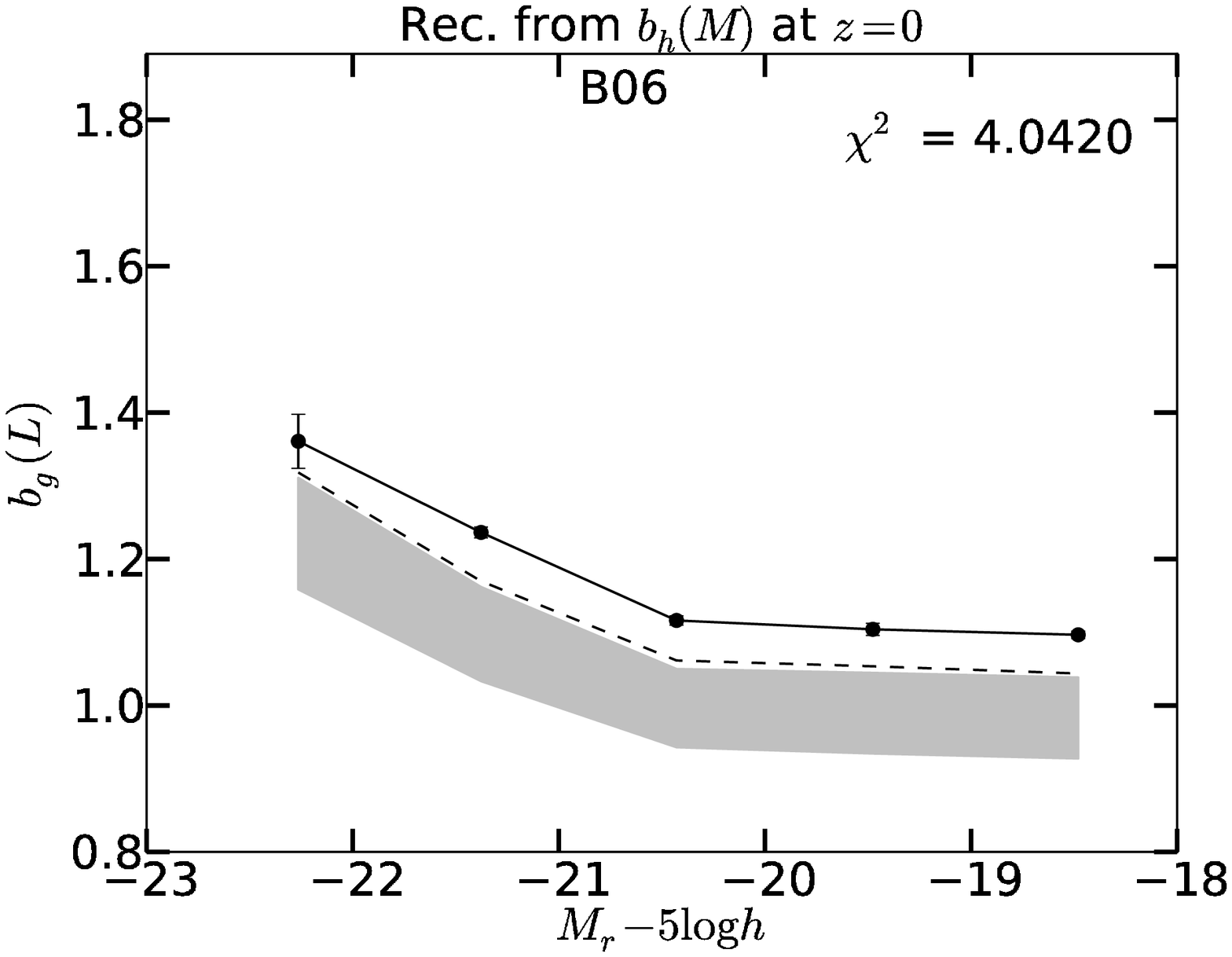}
\includegraphics[width=84mm]{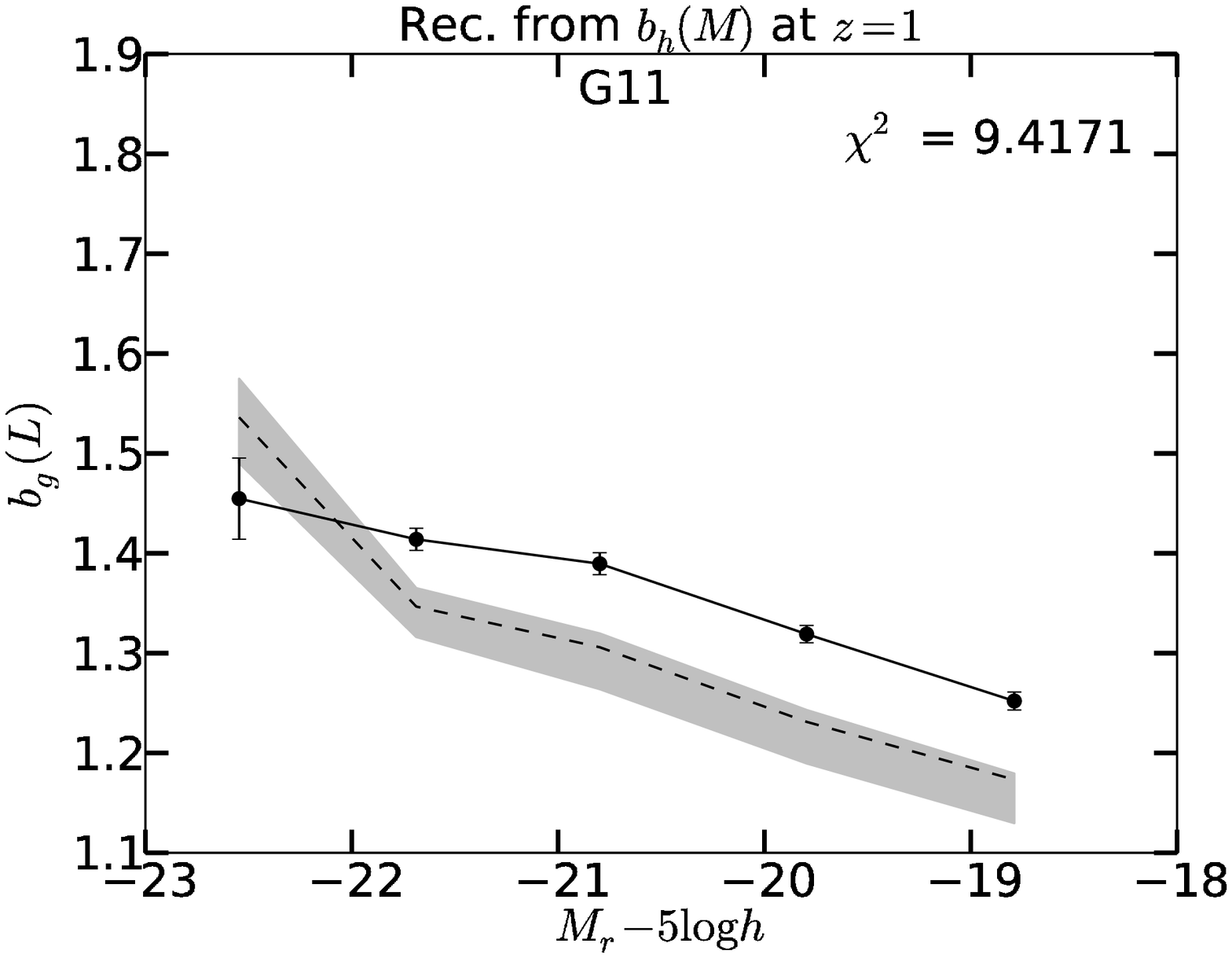}
\includegraphics[width=84mm]{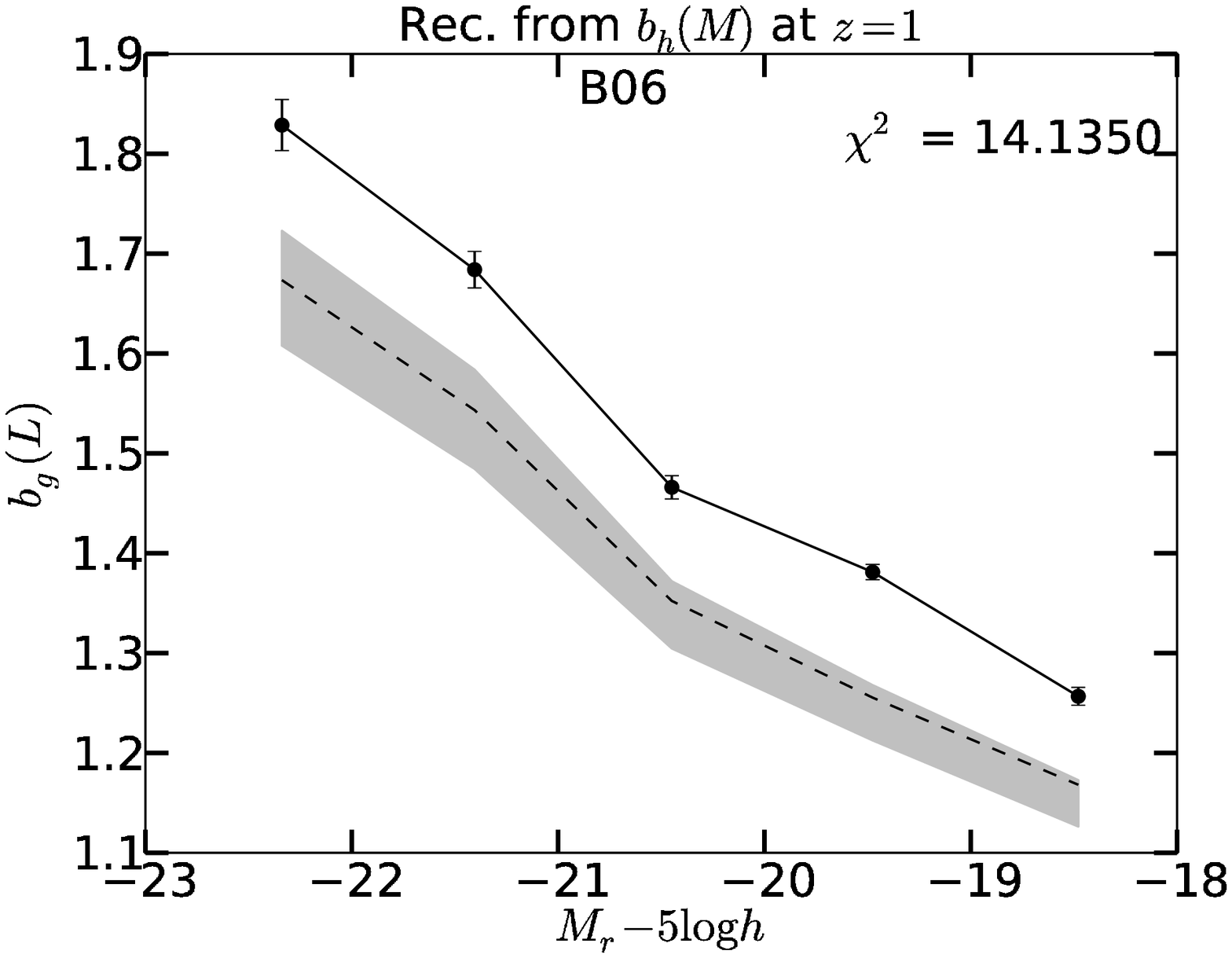}
\caption[Reconstruction of galaxy bias from main halo bias]{The same as in Fig.\ \ref{fig:rec_gal_fof}, but the reconstructions are obtained from $b_h(M)$ instead of $b_{FOF}(M)$. }
\label{fig:rec_gal_msh}
\par\end{centering}
\end{figure*}

In this section we want to measure if we can recover galaxy bias by assuming that the HOD and galaxy clustering depend only on halo mass. To do this we make a reconstruction of $b_g(L)$, that we will call $b_{rec}(L)$, from the measurements of $b_{FOF}(M)$ (or $b_h(M)$) and the occupation of galaxies in these haloes. If we are in the linear regime (as we are) and the occupation of galaxies is only halo mass dependent, then the value of $b_g(L)$ must coincide with the reconstruction $b_{rec}(L)$ obtained from the following expression (Scoccimarro \etal 2001, Cooray \etal 2004, Wechsler \etal 2006, Coupon \etal 2012):
\begin{equation}
b_{rec} (L) = \int dM b_{h} (M) n_{h} (M) \frac{\langle N_{g,h}(M,L) \rangle}{n_g(L)}
\label{eq:rec}
\end{equation}
where $n$ corresponds to the number density of the galaxies or haloes and $N_{g,h}(M,L)$ is the mean number of galaxies per halo (or FOF) of mass $M$. We will test these $b_{rec}(L)$ for both FOF and haloes. As we use a range in halo mass, the galaxy sample is restricted to those objects that are inside the considered haloes, excluding those which are outside the range of halo mass. Because the FOFs and haloes represent exactly the same objects, one could think that they should give the same results. But their definitions of mass are different, and these differences are important for large and unrelaxed haloes. For this, the mass dependencies of clustering and the relations between mass and galaxy occupation can be stronger for one definition than for the other, and the effects of large and unrelaxed haloes will produce differences in the reconstructions.

The error is obtained by calculating the Jack-Knife error of the reconstruction using $64$ cubic subsamples. Finally, in all the reconstructions $\chi ^2/\nu$ is calculated according to the formula:

\begin{equation}
\chi ^2/\nu = \frac{1}{N} \sum_i^N \frac{M_i - R_i}{\sigma_{M, i}^2 + \sigma_{R, i}^2}, 
\label{eq:chi2}
\end{equation}
where $N$ is the number of data points, $M$ and $R$ are the measured and reconstructed points respectively, and $\sigma_{M, i}$ and $\sigma_{R, i}$ are the respective errors of $M$ and $R$. We assume that the reconstructions of each subsample is independent of each other.

We will focus on the G11 and B06 models in $M_r$ as representatives of MPA and Durham models to analyse the reconstructions. Although the SAMs have clear differences in the luminosity dependence of $b_g(L)$, the results of the reconstructions of all the SAMs present similar behaviours and the same qualitative conclusions than in G11 and B06. 

Fig.\ \ref{fig:rec_gal_fof} shows the reconstructions of $b_g(L)$ from FOFs of these two models at $z = 0$ and $1$. The reconstructions $b_{rec}(L)$ are shown as a grey shaded region representing $b_{rec}(L) \pm 1 \sigma$, and they are compared to the real values, in solid lines. We also show as dashed lines $b_{rec}(L)$ using the Tinker \etal (2010) model for $b(M)$ instead of the measurements of the simulation in order to compare the differences between modelling and measuring $b_{FOF}(M)$ for the reconstruction. Top panels show $z = 0$ and bottom panels are at $z = 1$. Finally, in the left panels we see the reconstructions of G11, and in the right panels we used the B06 model. In Fig.\ \ref{fig:rec_gal_msh} we show the reconstructions of $b_g(L)$ from haloes instead of FOFs for the same SAMs and redshifts. 

First of all, we can see that the reconstructions using the Tinker \etal (2010) model differ with respect to the reconstructions from the measurements of $b_{FOF}(M)$ or $b_h(M)$ by the order of $1 \sigma$. The errors of the reconstructions reflect the fluctuations of the measurements of $b_{FOF}(M)$ and $b_h(M)$ in the simulation. Given the agreement between modelling and measuring bias from Fig.\ \ref{fig:hal_bias}, this difference in the reconstructions can be seen as a measurement of the effects of the fluctuations of $b_{FOF}(M)$ and $b_h(M)$ in the reconstruction. But, as the galaxies are located in the haloes of the simulation, these fluctuations in the bias should be included in the reconstruction if we want to study the relation between the galaxies and their haloes, so we focus in comparing $b_g(L)$ with $b_{rec}(L)$ from the measured halo bias (i.\ e.\ shaded region instead of dashed lines). 

Secondly, for the haloes we can see that the agreement between measurements and reconstructions tend to be better at low redshift, although for FOFs this is not so clear. On the other hand, the reconstructions tend to be different from the SAM measurements by a factor of $6-7\%$ at the level of $1\sigma$ for both FOFs and haloes. This differences in clustering corresponds to a $50\%$ difference in halo mass (see Fig.\ \ref{fig:hal_bias}). Another important result independent of the SAMs and redshift is the fact that FOFs predict better $ b_g(L) $ than haloes. This reflects that the unbinding processes, somehow, lose information about the galaxy clustering. In some sense, the mass of the FOF groups is more directly related to the clustering of these regions than the mass restricted to the bound particles in these overdensities, maybe because the FOF groups include more environment of the overdensity. Finally, we can see that the reconstructions tend to underpredict $b_g(L)$. This is a constant effect in $ z $ and appears using $b_{FOF}(M)$ or  $b_h(M)$ and for all the SAMs. This effect is analysed more in detail in \S \ref{sec:subhalo_pop} and \S \ref{sec:halo_mass}.

For equation (\ref{eq:rec}) to be accurate we need to satisfy on of the following to conditions. The first condition is that all the haloes of the same mass have the same clustering. If this is the case, all the galaxies in these haloes have the same clustering and then we are assigning the correct clustering for the galaxies. The second condition is that  galaxies populate haloes only according to their mass. If this is the case, even if the first condition is not satisfied the galaxies in the same masses must statistically have the same mean clustering. We have seen that the reconstructions differ from the measurement of $b_g(L)$, so this means that both conditions fail. So, for a fixed halo mass, different haloes must have different clustering (assembly bias). Moreover, the population of galaxies in haloes of the same mass must be correlated with the halo bias. In order to study this correlation, in \S \ref{sec:subhalo_pop} we will study the subhalo occupation dependence of bias. We will do this because we expect the number of subhaloes to be directly related with the number of galaxies but at the same time it is independent of the SAM. In \S \ref{sec:halo_mass} we will study the halo mass ranges where we see the assembly bias effects in the reconstructions.

\subsection{Subhalo population}\label{sec:subhalo_pop}

\begin{figure*}
\begin{centering}
\includegraphics[width=84mm]{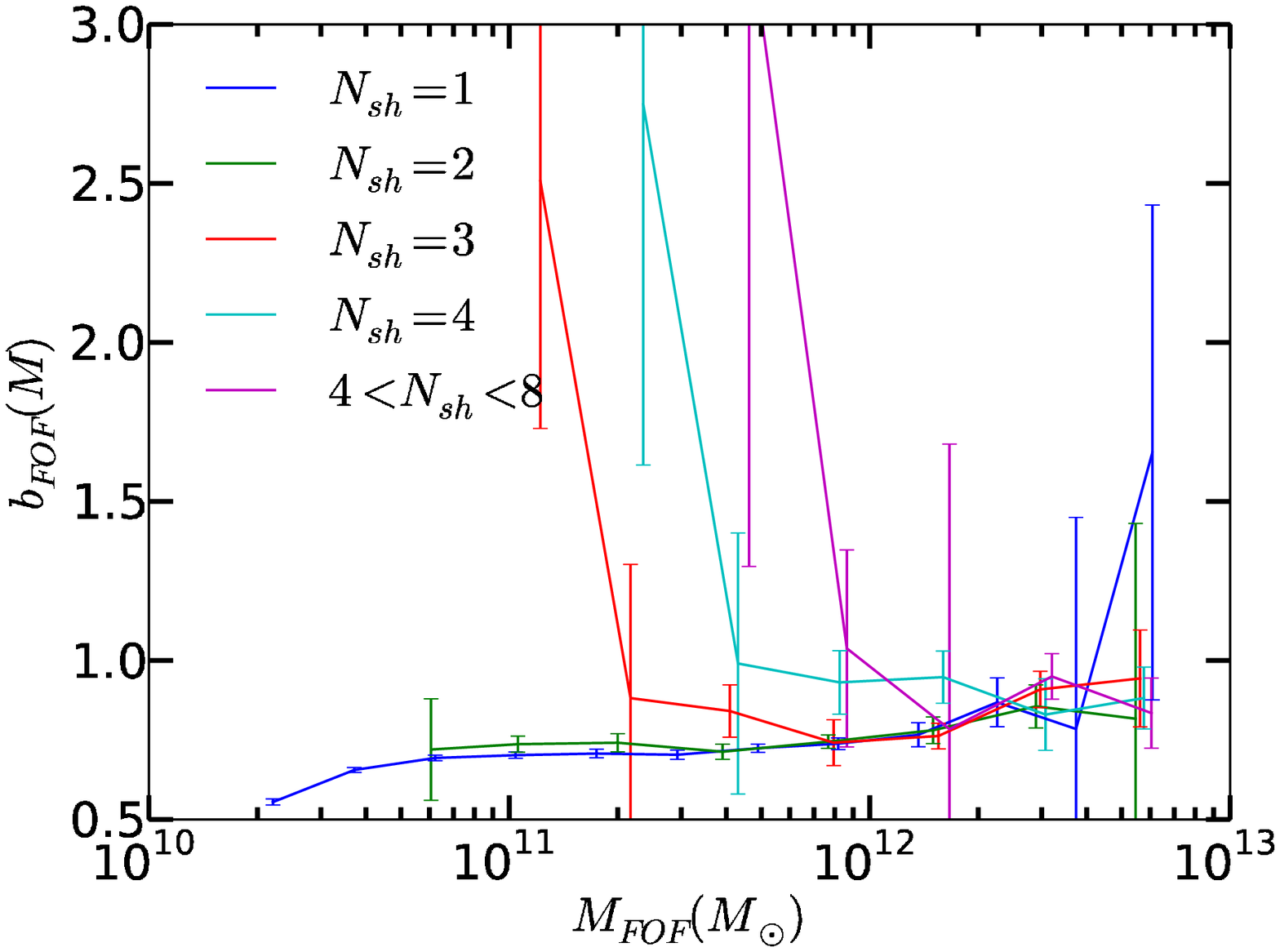}
\includegraphics[width=84mm]{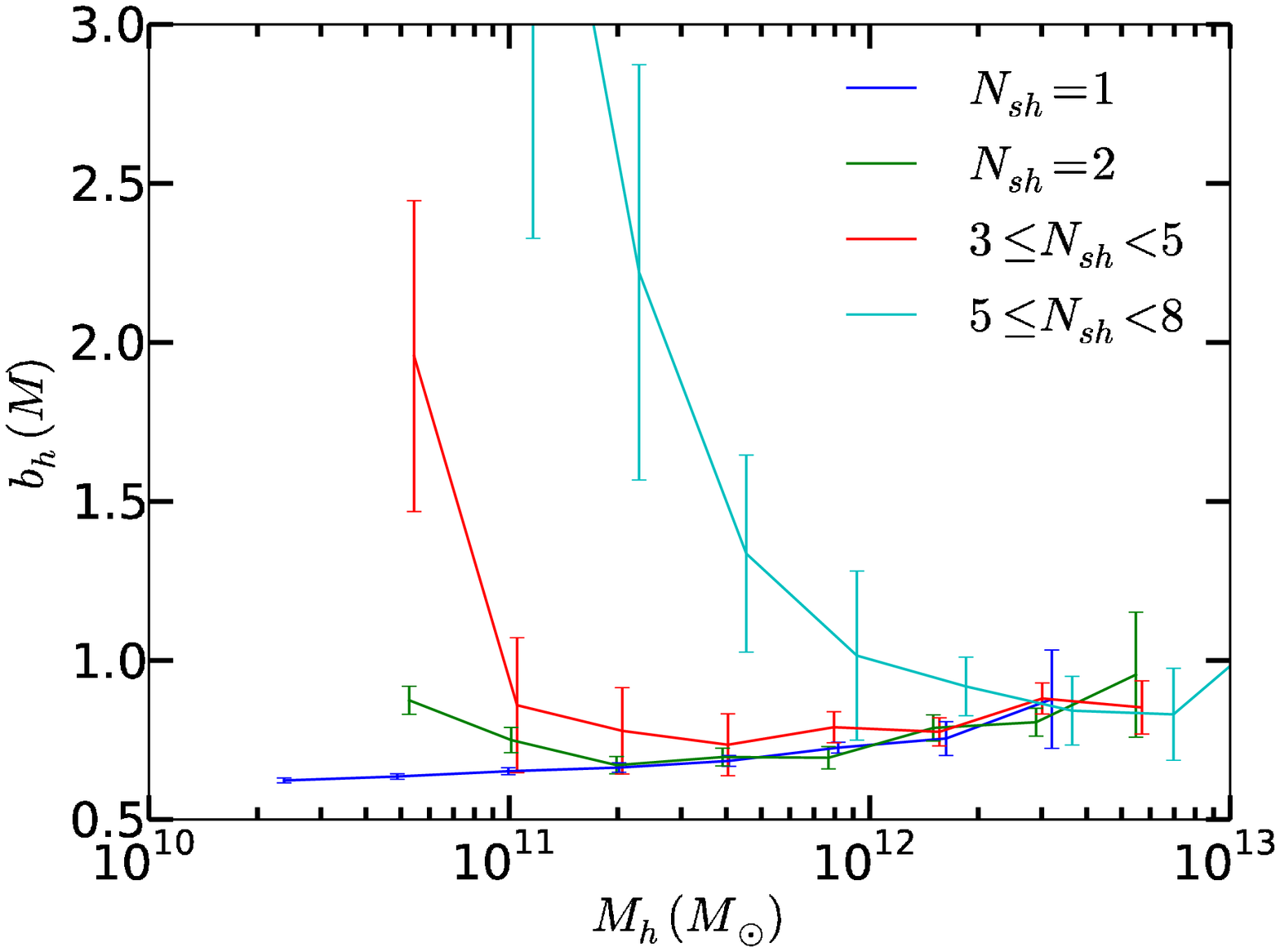}
\includegraphics[width=84mm]{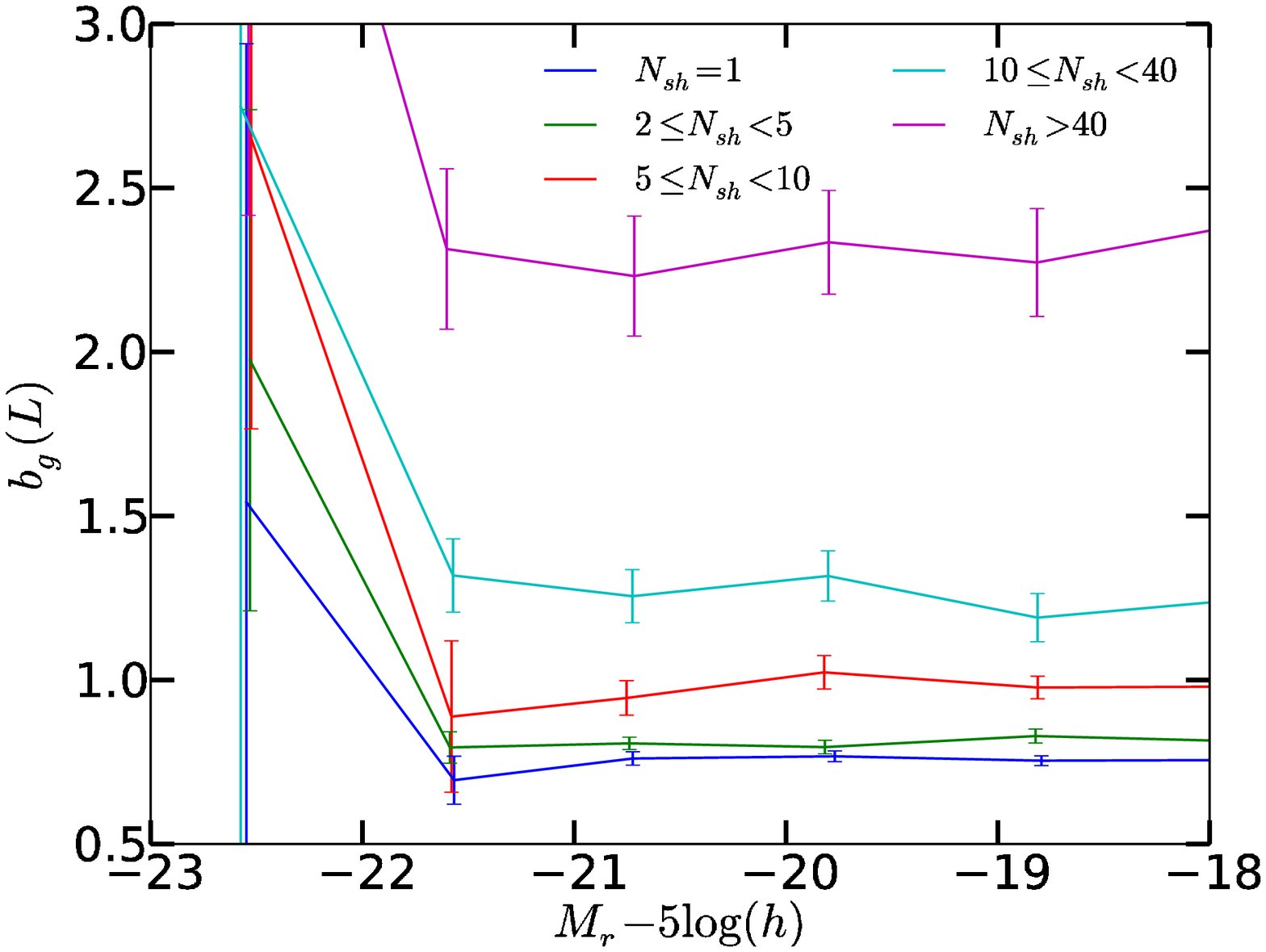}
\includegraphics[width=84mm]{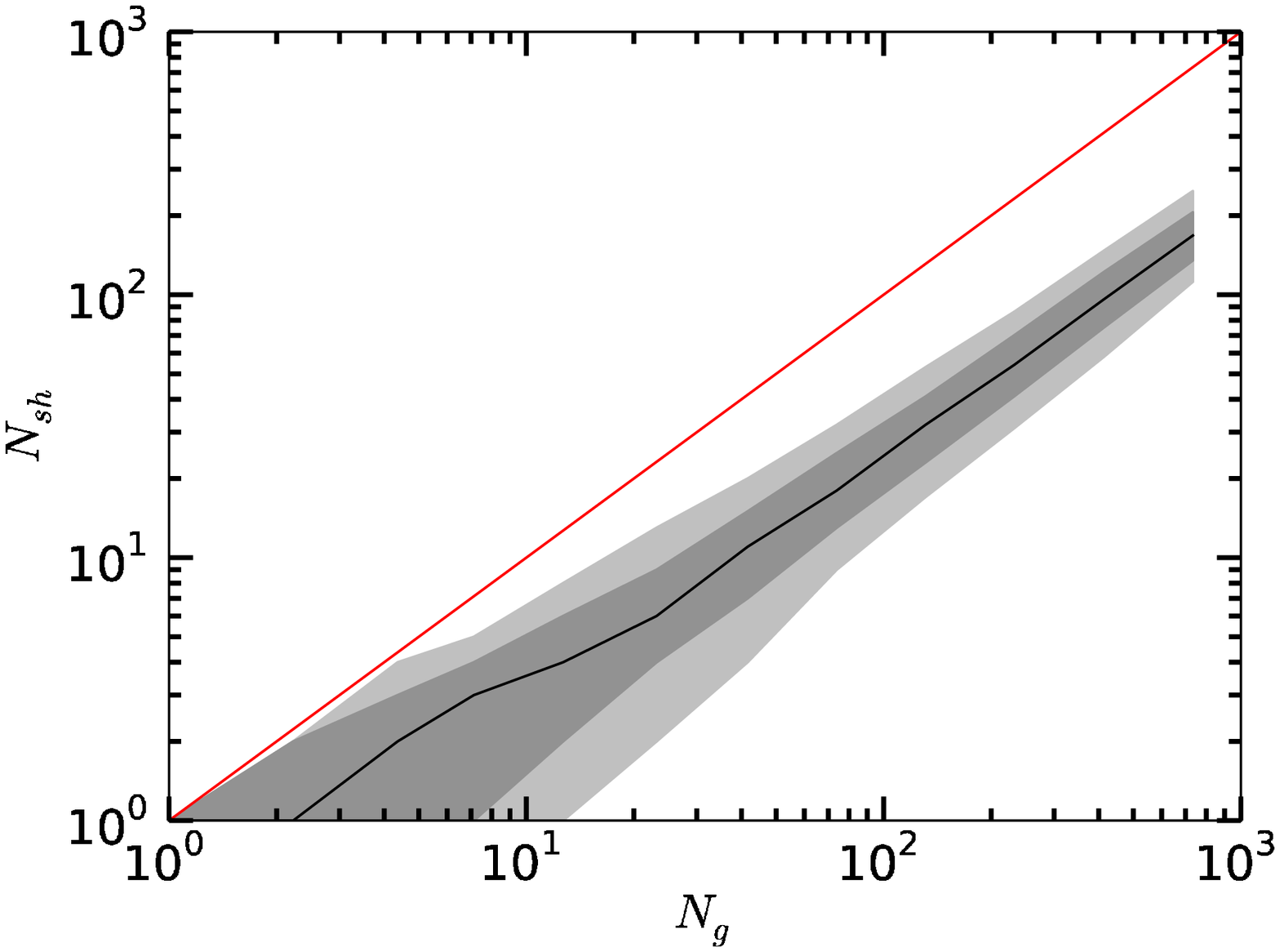}
\caption[$ b_{FOF}(M) $, $ b_{sh}(M) $ and $ b_g(L) $ for samples with different $ N_{sh}(M_h)$..]{Top: $b_{FOF}(M)$ (left) and $b_{h}(M)$ (right) for different samples according to their number of subhaloes $N_{sh}$ inside. Bottom left: $b(L)$ for galaxies in haloes with different $ N_{sh}$ for G11 model. Bottom right: number of galaxies ($N_g$) vs number of subhaloes ($N_{sh}$) of the FOFs. The grey shaded regions represent the $68$ and $95$ percentiles. The red line shows $N_g = N_{sh}$. All the panels are at $ z = 0 $.}
\label{fig:bias_N}
\par\end{centering}
\end{figure*}

In this section we study the subhalo occupation dependence of halo and galaxy bias. The idea is to separate the haloes and their content depending on the amount of subhaloes inside the haloes, $N_{sh}$, to see if clustering depends on their halo substructure.  This is an indirect measurement of environment, since the amount of subhaloes in a halo depends on the merging history of the halo, which is related to the environmental abundance of haloes. This is also interesting since the number of subhaloes in the haloes is related to the number of galaxies in it.

In Fig.\ \ref{fig:bias_N} we can see $ b_{FOF}(M) $ (top left), $ b_{h}(M) $ (top right) and $ b_g(L) $ (bottom left) separating the samples of FOFs, haloes and galaxies according to the number of subhaloes in the halo. The bottom right panel shows the relation between the number of galaxies ($N_g$) of the G11 model and the number of subhaloes ($N_{sh}$) in the FOFs. The red line represents $N_{sh} = N_g$. Two conclusions can be obtained from Fig.\ \ref{fig:bias_N} about galaxies and haloes: (1) for a fixed mass (at low masses at least) or luminosity, the dependence of $b$ on subhalo occupation, $N_{sh}$, is very strong, and (2) for a fixed $N_{sh}$ the dependence of $b_g(L)$ on $L$ is weak. So, galaxy clustering has a strong dependence on $N_{sh}$, meaning that for a fixed luminosity we are mixing galaxies with different clustering for the same reconstruction, and this can cause a deviation between the reconstruction and the real value of $b_g(L)$. Moreover, for a fixed mass, $b_{FOF}(M)$ and $b_h(M)$ present different clustering according to the number of subhaloes, while in the reconstruction we are wrongly assuming that bias only depends on mass. This is an indication of assembly bias. From bottom right panel we can see that the number of subhaloes increases with the number of galaxies. We can see that the haloes with a high number of subhaloes tend to have even more galaxies. As the haloes with higher $N_{sh}$ have more galaxies, the reconstructions produce an underestimation of $b_g(L)$, since we are assuming the same mean bias for these galaxies, while the haloes with more galaxies have a higher bias than the mean value of these masses. We used the number of subhaloes instead of the number of galaxies because it is independent of the SAM, and this is only dependent on the dark matter distribution, so we should see this effect in any simulated galaxy catalogue where the number of galaxies increases with the number of subhaloes. This is likely what should also happen in the real Universe. Note that this is not the case in HOD assuming  $P(N|M)$, by construction.

This can be seen as a galaxy clustering consequence of assembly bias, since we see that haloes of the same mass have different clustering, and we also see that this has consequences on the galaxy clustering predictions. This means that for a fixed mass the galaxies are not randomly distributed, so they depend on other properties than mass. If their distribution were random for each mass, then the reconstruction should work by definition. If we include $N_{sh}$ as another variable in the reconstructions, then the predictions of galaxy clustering are improved. We have checked this by reconstructing $b_g(L)$ and selecting the galaxies according to the number of subhaloes in their haloes. The results are significantly improved in this case (but they are not shown here).

From Fig.\ \ref{fig:bias_N} we can see that the subhalo occupation dependence of clustering is stronger for low masses. In order to see explicitly the masses where we see this effect, in \S \ref{sec:halo_mass} we study the mass dependence of the reconstructions.

\subsection{Halo mass dependence}\label{sec:halo_mass}

\begin{figure}
\begin{centering}
\includegraphics[width=84mm]{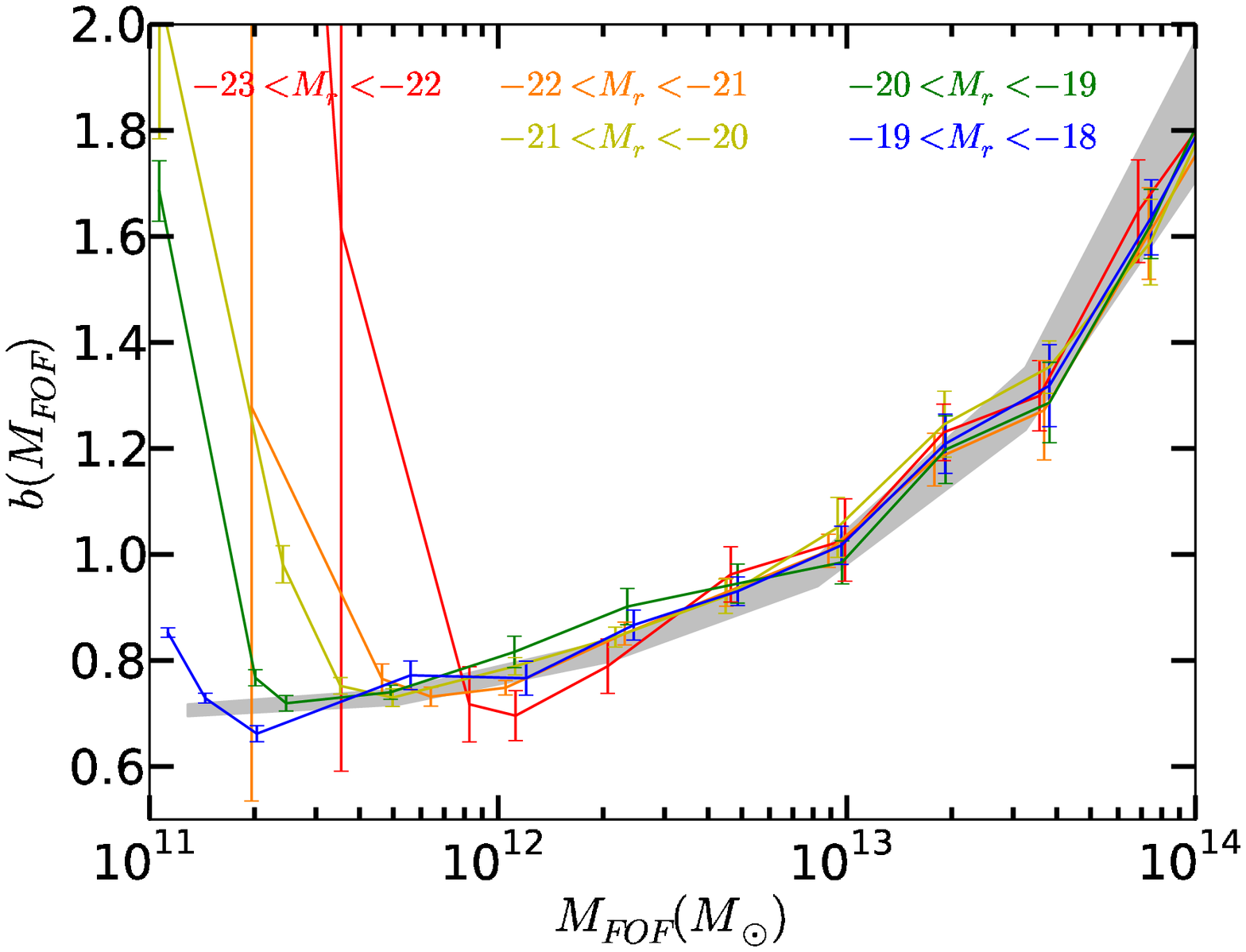}
\includegraphics[width=84mm]{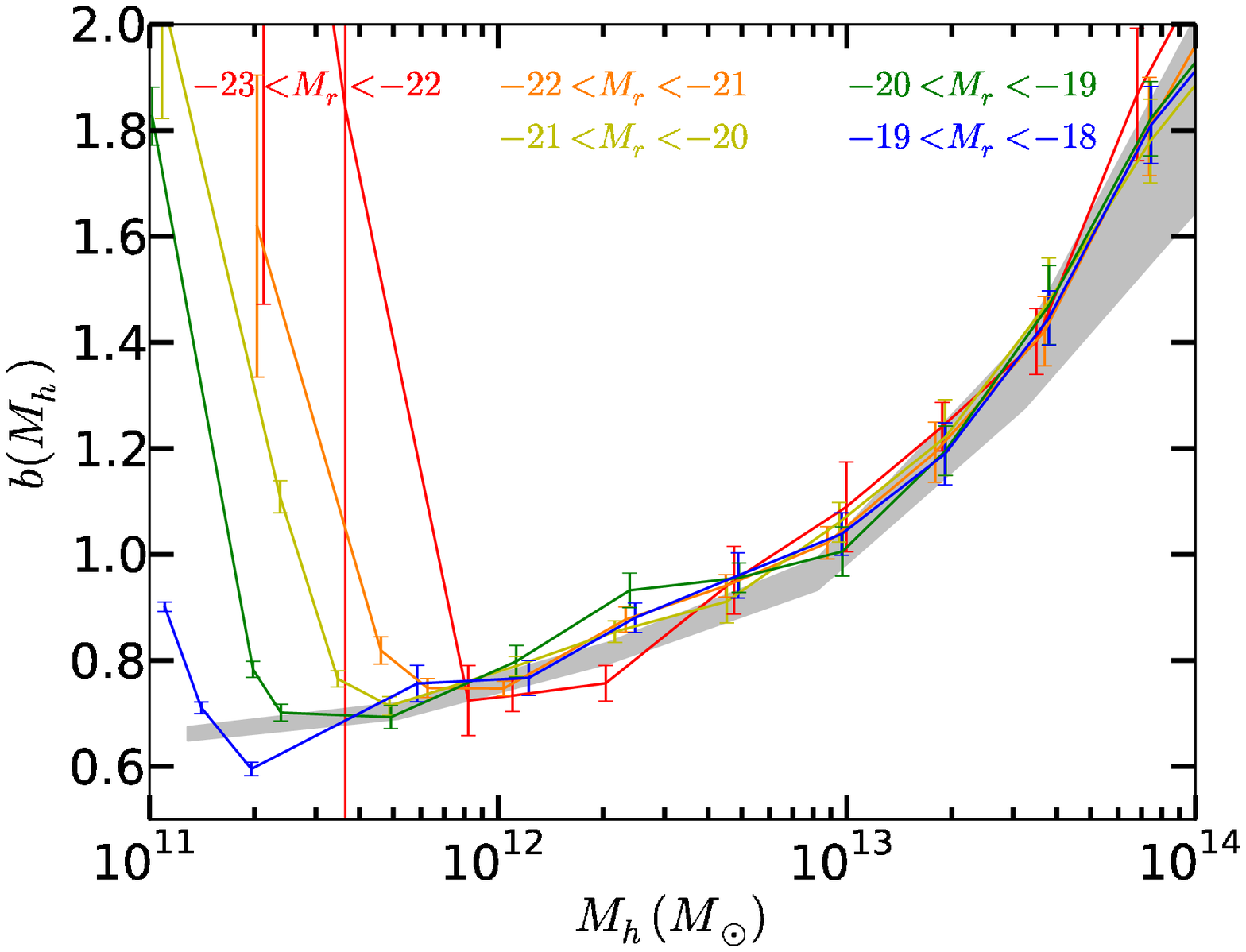}
\caption[$b_g(M)$ at different luminosity bins.]{$b_g(M)$ and $b_{rec}(M)$ (from FOFs on top and haloes on bottom) for different luminosity bins for G11 galaxies. Solid lines represent galaxy bias as a function of FOF mass (top) or halo mass (bottom) of their host haloes with different luminosities represented by different colours. The grey shaded zone refer to the range of $b_{rec}(M) \pm  1 \sigma$ from $b_{FOF}$ (top) and from $b_{h}$ (bottom). As the reconstructions are binned by halo mass, $b_{rec}(M) = b_{FOF}{M}$ (top) and $b_{rec}(M) = b_h(M)$ (bottom).}
\label{fig:bg_vs_np}
\par\end{centering}
\end{figure}

To study the mass dependence of the success in the reconstructions, we measured $ b_g(M_{FOF}) $ and $b_g(M_h)$ in several luminosity bins. Fig.\ \ref{fig:bg_vs_np} shows $b_g(M_{FOF})$ (top) and $b_g(M_h)$ (bottom) for galaxies at different luminosities. Each colour corresponds to a luminosity. The lines show $b_g(M)$ for galaxies in FOFs or haloes with the respective luminosity, and the grey shaded regions represent the measured ranges $b_{rec}(M) \pm  1 \sigma$ obtained from FOFs (top) and from haloes (bottom). This figure, then, allows us to see explicitly how the reconstructions of galaxy bias work at different masses. We have only used one mass bin for each reconstruction. For a narrow mass bin, the HOD reconstruction prediction equals the halo model value, so the shaded predictions in Fig.\ \ref{fig:bg_vs_np} equal those of Fig.\ \ref{fig:hal_bias}. Then, the reconstruction works if the values of $b_g(M)$ are close to $b_{FOF}(M)$ (and the same for $b_h(M)$). In Fig.\ \ref{fig:bg_vs_np} we can see two different behaviours. At high masses, $b_g(M)$ is close to $b_{FOF}(M)$ and $b_h(M)$, since the solid lines tend to be close or inside their shaded zones. This means that the reconstruction of $b_g(L)$ at these masses works, and the halo mass gives sufficient information to predict galaxy clustering. However, we note that there is an underestimation of $b_{rec}$ of the order of $1 \sigma$ when haloes are used instead of FOFs. In the low mass regions there is a strong disagreement between $b_g(M)$ and $b_{FOF}(M)$, and also with $b_h(M)$, especially for the brightest galaxies. The bias of the brightest galaxies is much higher than the mean one of the haloes of the corresponding mass. So, in these low masses, the galaxies are populated precisely in a way that the brightest galaxies are in the most clustered haloes of the corresponding mass. This means, again, that the clustering of these galaxies does not only depend on mass, and this is also another indication of assembly bias, since haloes of the same mass must have a different clustering. We also notice that the disagreement between $b_g(M)$ and $b_h(M)$ tends to be stronger than between $b_g(M)$ and $b_{FOF}(M)$. This is another indication that $M_{FOF}$ is more strongly related to galaxy clustering than $M_h$. This means that when the haloes and FOFs present important differences in their masses, $M_{FOF}$ tends to reflect better $b_g(L)$ than $M_h$.

We have seen that the distortions in the reconstructions appear at $M_h \lesssim 3-5 \times 10^{11} M_\odot$. Then, the reconstructions of $b_g(L)$ when we exclude haloes of these low masses and their galaxies work, and the predictions of galaxy clustering are correct. This low mass problem can be due to different aspects or a combination of them. First of all, it can reflect the consequences of assembly bias on galaxy clustering. Secondly, it can be affected by the strong stripping and mass distortions of haloes of the Millennium Simulation. When haloes interact with others or pass through high density environments, sometimes the masses are artificially distorted, and this effect is stronger for lower masses. Finally, the SAMs could be affected by assembly bias more strongly than reality. If this is the case, the clustering of SAMs at these masses would not be correct and we would need to exclude from the analysis those galaxies that reside in low mass haloes, regardless of their properties. But if we exclude these galaxies, then $b_g(L)$ is distorted and an excess of clustering for galaxies of $M_r > -20$ is found, meaning that we need to include these galaxies for clustering studies if we want to recover observations.

\section{Summary and discussion}\label{sec:conclusions}

We used the Millennium Simulation to study the clustering of galaxies as predicted by semi-analytical models (SAM) of galaxy formation and the dependence of clustering on luminosity in the SDSS $r$ band filter. We measured the clustering of haloes and we found good agreement with theoretical models, specially for the Tinker \etal (2010) model. We have found discrepancies in the galaxy clustering with respect to observations (Zehavi \etal 2011) that can be due in part to their excess of bright galaxies and in part due to the differences on the assumed cosmology for the analysis, in particular the difference in $\sigma_8$. Although the SAMs are not based on the HOD model, their populations agree with the observations of the HOD from the SDSS DR7 (Zehavi \etal 2011). We try to reproduce galaxy bias from the bias of FOF groups and gravitationally bound haloes as a function of mass, by assuming that the population of galaxies only depends on the mass of these objects. From our study we obtain the following results:

(i) Although in some cases the reconstructions can provide a good $\chi^2/\nu$, the reconstructions tend to underpredict $b_g(L)$ by a factor of $\gtrsim 5 \%$. This translates to an error in the inferred halo mass of the order of  $50\%$.

(ii) FOF groups make better reconstructions of $b_g(L)$ than haloes, specially at high redshift, which could be due to the fact that FOF groups include more information about the environment than the main haloes. 

(iii) The clustering of haloes and galaxies depends strongly on the amount of substructure in their host haloes. For a fixed halo occupation (of subhaloes), the luminosity dependence of $b_g$ is very weak. For a fixed halo mass, there is a strong dependence of $b_h$ on the occupation of subhaloes, an indication of assembly bias. This result is independent of the SAMs. 

(iv) The reconstructions of $b_g$ from haloes work better at high masses, but some diagreements with $b_g$ come from the low mass haloes, where the assembly bias effect is stronger. This effect is stronger for haloes than for FOFs. This means that when the masses of haloes and FOFs are significantly different, the mass of the FOFs reflects better $b_g(L)$ than the mass of the haloes. This effect occurs for $M_h \lesssim 3-5 \times 10^{11} M_\odot$. The suppression of galaxies in the smallest haloes avoids the problems of the reconstructions of $b_g(L)$, but changes the shape of the luminosity dependence of $b_g(L)$, which makes it inconsistent with observations from SDSS DR7 (Zehavi \etal 2011).

Our results can also depend on the halo and subhalo finders and the SAMs. On one hand, subhaloes in the Millennium Simulation suffer very strong stripping when they interact with high density environments, and this can have consequences on the assembly bias found at low masses. On the other hand, SAMs have been modelled from these dark matter objects, and their clustering consequences could also present artificial dependences on assembly bias. A galaxy catalog constructed from an HOD model only using halo mass would not reflect this effect. However, we expect to find this effect for all the galaxy formation models where the number of galaxies increases with the number of subhaloes, and also in the real Universe.

Recent studies (Tissera \etal 2010, Sawala \etal 2012) indicate that we need to take into account baryonic effects on the dark matter haloes. The density profile of haloes change substantially when baryons are included, and this change can produce important effects on the galaxy formation models applied (Tissera \etal 2010). Sawala \etal (2012) also saw that baryonic physics reduces the mass and abundance of haloes below $M_h < 10^{12}$. Our study indicates that it would be premature to use the HOD interpretation for such halo masses to study these baryonic effects using the clustering of galaxies. 

The agreement of Tinker \etal (2010) model, together with the convergence of Fig.\ \ref {fig:bg_vs_np} at the largest masses studied, seem to indicate that the SAMs agree with our assumptions at large masses. However, other dependences than mass are needed to predict the clustering of the SAMs on small masses, then care must be taken when assuming the HOD model at masses below $3-5 \times 10^{11} M_\odot$, especially when assuming that galaxy clustering only depends on mass. We have seen that the halo clustering depends strongly on $N_{sh}$ for a fixed halo mass, and this explains the discrepancies between the measurements in SAM and the HOD modelling. For any galaxy formation model where $N_g \varpropto N_{sh}$ we would expect a similar assembly bias. In this case, we will underestimate the galaxy bias with the HOD modelling. 

Frequently, the HOD is assumed to relate galaxy properties and halo masses (Zehavi \etal 2011, Coupon \etal 2012, Cooray \etal 2012). In the case of Zehavi \etal (2011), they measure the HOD parameters form the clustering of galaxies assuming that the clustering only depends on the halo mass. But these results can be affected by the halo bias dependence on the subhalo occupation for fixed masses if the number of galaxies increases with the number of subhaloes. This conclusion seems quite generic in the light of our Fig.\ \ref{fig:bias_N}: for a fix halo mass, clustering is stronger for halos with more substructure.  
The standard implementation of the HOD assumes that the clustering of galaxies only depends on the mass of the haloes.
If for a fixed halo mass there are more galaxies in the halos with more subhalos, then the mean clustering of these galaxies
will be higher than the mean clustering of the haloes and we will wrongly conclude that they are in more massive haloes. This results in
an overestimation of the halo mass using clustering. We have shown that this is the case in SAM, but we expect this to also
be true in any other model of galaxy formation where the number of galaxies correlate with the number of subhalos.

\begin{figure}
\begin{centering}
\includegraphics[width=84mm]{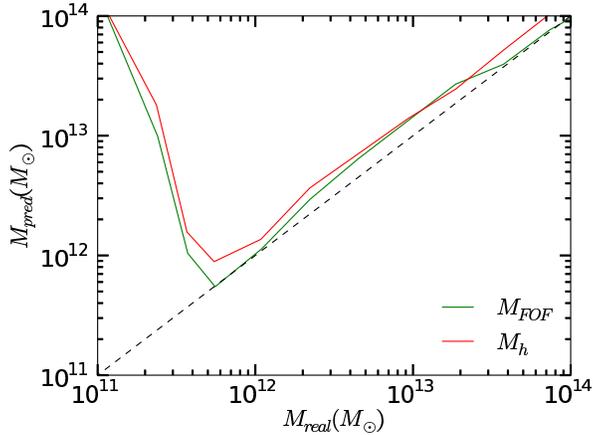}
\caption[Halo mass prediction compared to real halo mass.]{Comparison between the halo mass of the host haloes of galaxies with $M_r < -20$ with the predicted halo mass obtained from the galaxy clustering. This can be obtained from figure \ref{fig:bg_vs_np} by  translating the bias to halo mass for galaxies with $M_r<20$. In green we used FOF, while in red we show the predictions for haloes. The dashed line shows $M_{pred} = M_{real}$.}
\label{fig:mpred_vs_mtrue}
\par\end{centering}
\end{figure}

In order to do a simple estimation of the order of magnitude of this overprediction, in Figure \ref{fig:mpred_vs_mtrue} we show the relation between the real mass of the haloes and the mass predicted from the clustering of their galaxies. As the mean clustering of the galaxies is higher than the mean clustering of their haloes, the galaxy bias corresponds to the bias of a higher halo mass. In this figure we see the overprediction for galaxies brighter than $M_r < -20$. 

We note the degeneracy on the predicted mass over the real mass. We can see for example that a predicted mass of $10^{13} M_\odot$ corresponds to both a real mass of $2-3 \times 10^{11}$ and a real mass of $7-8 \times 10^{12}$. This means that we cannot strictly predict the mass from the clustering amplitude only, since this degeneracy affects all the masses.  In particular we will never get the true mass for the lower mass halos.

From Table $3$ of Zehavi \etal 2011 we see that the best HOD parameters obtained are $\log M_{min} = 11.83$, $\log M_0 = 12.35$ and $\log M_1 = 13.08$. If the relation of Figure \ref{fig:mpred_vs_mtrue} were correct, the parameters should be corrected to $\log M_{min} \simeq 11.74-11.83$, $\log M_0 \simeq 12.18-12.25$ and $\log M_1 \simeq 12.90-12.92$. For this rough estimation we ignore the degeneracy and choose the closest $M_{pred}$, since for these masses most of the galaxies are populating large haloes.  In detail we should weight the prediction by the number of objects. Similar considerations could be applied to other HOD analysis (Coupon \etal 2012, Cooray \etal 2012) where assembly could play a role, specially for the lower masses.

\section*{Acknowledgements}

We thank Marc Manera for interesting discussions and useful ideas. We also thank Carlton Baugh, Sergio Contreras, Qi Guo, Noelia Jim\'{e}nez, Cedric Lacey and Ravi Sheth for their comments and discussions about this project. A.P.\ wants to thank also Albert Izard for useful code support. The Millennium Simulation databases used in this paper and the web application providing online access to them were constructed as part of the activities of the German Astrophysical Virtual Observatory (GAVO). Funding for this project was partially provided by the Spanish Ministerio de Ciencia e Innovacion (MICINN), project AYA2009-13936, Consolider-Ingenio CSD2007- 00060, European Commission Marie Curie Initial Training Network CosmoComp (PITN-GA-2009-238356), research project 2009- SGR-1398 from Generalitat de Catalunya. A.P. was supported by beca FI from Generalitat de Catalunya. We acknowledge support from the European Commission's Framework Programme 7, through the Marie Curie International Research Staff Exchange Scheme LACEGAL (PIRSES-GA-2010-269264)


\end{document}